\documentclass[oupdraft]{bio}
\usepackage[colorlinks=true, urlcolor=citecolor, linkcolor=citecolor, citecolor=citecolor]{hyperref}

\usepackage{fixltx2e}
\usepackage{float}

  \usepackage{breqn}

\usepackage{bm,graphicx,color,amsbsy,amsmath,placeins,
  amsthm,amsfonts,amssymb,setspace,lscape}
\usepackage{caption}
\usepackage{tabularx}

  \def\bSig\mathbf{\Sigma}

\usepackage{amsfonts}
\usepackage{amsmath}
\usepackage[FIGTOPCAP]{subfigure}
\usepackage{url}

\newcommand{\Mb}{\mathbf{M}}

\newcommand{\bM}{\bm{M}}

\newcommand{\bP}{\bm{P}}

\newcommand{\bX}{\bm{X}}

\newcommand{\cD}{\mathcal{D}}

\newcommand{\cM}{\mathcal{M}}

\newcommand{\EE}{\mathbb{E}}
\newcommand{\VV}{\mathbb{V}}

\newcommand{\PP}{\mathbb{P}}

\newcommand{\btheta}{\bm{\theta}}

\newcommand{\bnu}{\bm{\nu}}

\newcommand{\bpi}{\bm{\pi}}

\newcommand{\bpsi}{\bm{\psi}}
\newcommand{\bomega}{\bm{\omega}}

\newcommand{\bTheta}{\bm{\Theta}}

\newcommand{\bPsi}{\bm{\Psi}}
\newcommand{\bOmega}{\bm{\Omega}}

\def\bEta{\bm{\eta}}

\def\P{{\mathbb P}}

\def\logit{{\sf logit}}
  \usepackage{circuitikz}
\newcommand{\filleddiode}{\ctikzset{bipoles/length=.7cm}\begin{circuitikz}[scale=.4] \draw
  (0,0) to[full diode] (2,0)
  ; \end{circuitikz}}

    \allowdisplaybreaks

      \newcommand*\mean[1]{\overline{#1}}
        \newcommand{\superscript}[1]{\ensuremath{^{\textrm{#1}}}}

          \newcommand{\msd}[2]{{#1}{\footnotesize({#2})}}
            \newcommand{\marg}{\superscript{\sf \footnotesize M}}
            \newcommand{\pname}[1]{{\small \sf{#1}}}

              \newenvironment{myindentpar}[1]%
              {\begin{list}{}%
                {\setlength{\leftmargin}{#1}}%
                  \item[]%
                }
                  {\end{list}}

\begin{document}
                  
                  \title{Nested Partially-Latent Class Models for Dependent Binary Data; Estimating Disease Etiology}
                  
                  \author{ZHENKE WU$^{\ast,1}$, MARIA DELORIA-KNOLL$^2$, SCOTT L. ZEGER$^1$\\[4pt]
                    \textit{$^1$ Department of Biostatistics, Johns Hopkins University, Baltimore, MD 21205\\}
                    \textit{$^2$ Department of International Health, Johns Hopkins University, Baltimore, MD 21205}
                    \\[2pt]
                    {zhwu@jhu.edu}}

                    \markboth%
                  {Z. Wu and others}
                  {Nested Partially-Latent Class Models for Dependent Data}
                  
                  \maketitle
                  
                  \footnotetext{To whom correspondence should be addressed.}

                  \begin{abstract}
                  {The Pneumonia Etiology Research for Child Health (PERCH) study seeks to use modern measurement technology to infer the causes of pneumonia for which gold-standard evidence is unavailable. The paper describes a latent variable model designed to infer from case-control data the etiology distribution for the population of cases, and for an individual case given his or her measurements. We assume each observation is drawn from a mixture model for which each component represents one cause or disease class. The model addresses a major limitation of the traditional latent class approach by taking account of residual dependence among multivariate binary outcome given disease class,  hence reduces estimation bias, retains efficiency and offers more valid inference. Such ``local dependence" on a single subject is induced in the model by nesting latent subclasses within each disease class. Measurement precision and covariation can be estimated using the control sample for whom the class is known. In a Bayesian framework, we use stick-breaking priors on the subclass indicators for model-averaged inference across different numbers of subclasses. Assessment of model fit and individual diagnosis are done using posterior samples drawn by Gibbs sampling. We demonstrate the utility of the method on simulated and on the motivating PERCH data. }
                  {Bayesian methods; Case-control studies; Local dependence; Latent class model; Measurement error; Disease etiology.
                  }
                  \end{abstract}
                  
                  \section{Introduction}
                  Clinicians routinely use measurements to differentially diagnose a patient's unknown disease etiology and then choose a treatment from among those available. More often than not, the differential diagnosis is a qualitative process based on judgment and experience. As clinical measurements become more precise and complex and as the number of possible known etiologies grows, such qualitative processes are less likely to be optimal. An important question therefore is whether formal probabalistic calculations can improve clinical decisions when the relevant information is quantitative. For example, in the Pneumonia Etiology Research for Child Health (PERCH) study of childhood pneumonia \citep{Levine2012}, a vector of presence/absence indicators for a large number of pathogens is measured on each child by polymerase chain reaction (PCR) using specimens from the nasopharyngeal cavity. A clinical goal is to use the multivariate binary response to infer the pathogen in the child's lung causing pneumonia.
                  
                  In addition, public health researchers are interested in estimating the population fraction of cases caused by each pathogen, referred to as the \textit{etiologic fractions} or \textit{population etiology distribution} \citep{feikin2014use}. Knowledge of the etiology distribution is essential for planning prevention and treatment programs. Because the lung cannot be directly sampled, except in cases of critical illness, imperfect measurements from the periphery are used to infer the \textit{latent state} of the disease. 
                  
\begin{figure}[H]
                                                                                                                                                                                                                                                                                                                                                                                                                                                                                                                                                                                                                                                                                                                                                                                                                                                                                                                                                                                                                                                                              \begin{center}
                                                                                                                                                                                                                                                                                                                                                                                                                                                                                                                                                                                                                                                                                                                                                                                                                                                                                                                                                                                                                                                                              \includegraphics[width=0.8\linewidth]{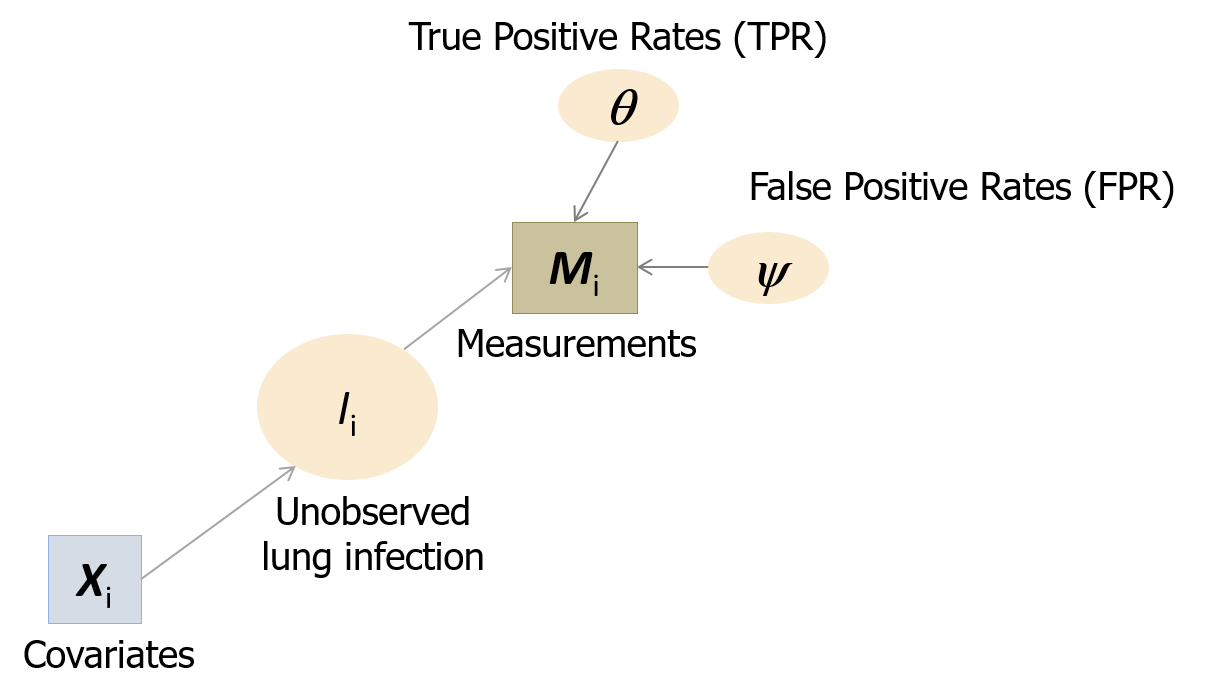}
                                                                                                                                                                                                                                                                                                                                                                                                                                                                                                                                                                                                                                                                                                                                                                                                                                                                                                                                                                                                                                                                              \end{center}
                                                                                                                                                                                                                                                                                                                                                                                                                                                                                                                                                                                                                                                                                                                                                                                                                                                                                                                                                                                                                                                                              \caption{Schematic of PERCH problem for an individual case $i$. $I_i$ for latent state; $\bM_i$ for multivariate binary measurements; $\bTheta$ and $\bPsi$ for true- and false-positive rates; $\bX_i$ for covariates. Quantities in rectangles are observed and those in ovals are unknown.}
                                                                                                                                                                                                                                                                                                                                                                                                                                                                                                                                                                                                                                                                                                                                                                                                                                                                                                                                                                                                                                                                              \label{fig:measurement_for_individual_i}
                                                                                                                                                                                                                                                                                                                                                                                                                                                                                                                                                                                                                                                                                                                                                                                                                                                                                                                                                                                                                                                                              \end{figure}

                  Figure \ref{fig:measurement_for_individual_i} summarizes the relations among measurements, covariates and lung infection for an individual case. PERCH intends to infer her latent  lung infection status ($I_i$, the latent state) by collecting multivariate binary measurements $\bM_i$ from the periphery. The joint distribution for $\bM_i$ is characterized by the true- and false- positive rates and the distribution of the latent disease-causing infection. Covariates such as age and HIV status can also influence the chance for each pathogen causing her disease. 
                  
                  In general terms, the PERCH scientific questions require inference about latent random variables \citep[e.g.][]{bollen2002latent}. The same is true for many other problems, for example, biomarkers for disease diagnosis \citep[e.g.][]{jokinen2010estimating}, words for learning topics of a text \cite[e.g.][]{hofmann2001unsupervised}, and questionnaire items for evaluating severity of depression \citep[e.g.][]{kroenke2002phq}. One way of classifying latent variable models is by the discrete or continuous nature of their latent and manifest (observed) variables. Among them, ``latent class" models (LCM) for discrete latent and discrete manifest variables were developed and widely applied since the 1950s \citep[e.g.][]{lazarsfeld1950thelogical, anderson1954estimation, lazarsfeld1968latent, Goodman1974}.
                  
                  LCMs constitute a family of distributions for correlated discrete measurements. The conventional LCM generally makes \textit{local independence} (LI) assumption that manifest variables are independent of one another given the latent class \citep{lord1952relation, lazarsfeld1959latent,  mcdonald1981dimensionality, bartholomew2011latent}. In the multivariate binary case, individual $i$'s measurement vector, $\bm{M}_i  = (M_{i1}, ..., M_{iJ})'$, is linked to her latent class ($I_i$) by the simple product likelihood $\P(\bM_{i} \mid I_i  = \ell, \bm{\theta})= \prod_{j=1}^J \P(M_{ij} \mid I_i = \ell, \bm{\theta})$, where ${\bm{\theta}}$ represents the collection of measurement parameters --- sensitivities and specificities. We then obtain the observed likelihood by summing over all the possible values of $I_i$, i.e., $\P(\bM_i  \mid \bm{\theta}, \bm{\pi}) = \sum_{\ell=1}^L \pi_{\ell} \prod_{j=1}^J \P(M_{ij} \mid I_i=\ell, \bm{\theta})$, where ${\bpi}$ is a vector of mixing weights of length $L$. The LI assumption implies that the latent memberships $I_i$ completely explains the marginal dependence in $\bm{M}_i$. Under local identifiability conditions \citep{Jones2010}, we can estimate $\bpi$ and $\btheta$ by the values that optimally reduce the observed dependence among measurements given latent class. The expectation-maximization (EM) algorithm \citep{Goodman1974} is one popular approach. Individual classification can then proceed by applying Bayes rule using the estimated parameters. 
                  
                  When classes are observed for some subjects, for example, motivated by the known control infection status $I_i=0$, \cite{wu2015} introduced a ``partially-latent" class model (pLCM). The control sample provides the requisite information to estimate the specificities of the measurements. In the original formulation, they assumed LI for the multivariate binary measurements within each class. However, within cases or controls, several pairs of pathogens had observed log odds ratios that are inconsistent with their model-based predictive distributions. To address this lack of fit in the covariances, one approach is to extend pLCM by introducing dependence among measurements for persons within the same class. These associations have scientific value in their own right, for example, to study patterns of pathogen-pathogen stimulation or inhibition.
                  
                  Deviations from LI, or ``local dependence" (LD) can occur in many applications, for example, in medical diagnostic tests when most severely diseased patients and the healthiest patients are easiest to correctly classify \citep{albert2001latent}, or when tests target on similar genetic molecules \citep{Qu1998}. Many authors have noted that not accounting for LD can bias estimates of model parameters \citep[e.g.][]{vacek1985effect, torrance1997effects, pepe2007insights}. Therefore, in many applications where the LI model for $[\bm{M}_i\mid I_i]$ is assumed, model adequacy is studied to ensure valid model-based conclusions \citep[e.g.][]{garrett2000latent, wu2015}.
                  
                  Ideas for relaxing LI can be distinguished by whether or not extra latent variables are introduced. Without doing so, \cite{harper1972local} modeled associations between pairs, triples, and higher order combinations of variables given latent class. \citet{haberman1979analysis} and \cite{Espeland1989} used log-linear models to extend LCM viewing the latent class as one of the category variables.  See also \cite{hagenaars1988latent} and \cite{yang1997latent}.
                  
                  The second approach allows for dependence by using extra latent variables of continuous or discrete types or a mixture. For example,
                  \cite{Qu1996} used Gaussian random intercepts to induce within-subject symmetric and positive correlations among multiple diagnostic tests. \cite{xu2009probit} used probit latent class models for more complex LD structures. \citet{albert2001latent} proposed to nest one extra unobserved subclass within each of two latent classes (diseased or non-diseased) to represent subjects measured without error. \citet{dendukuri2009modeling} hierarchically layered extra mixed latent variables in a Bayesian framework. Adding extra latent variables can account for LD because any multivariate discrete distribution can be represented by a locally independent LCM with sufficiently many latent classes \citep[Corollary 1]{dunson2009nonparametric}. However, when a satisfactory fit requires many classes --- especially when the dimension of manifest variables is high --- interpreting inferred classes remains a difficult task. 
                  
                  In this paper, we build on the second strategy and develop a novel latent variable model for multivariate binary data obtained from a \textit{case-control} study. Using control data with a known class and assuming the covariation among control measurements is shared among the other latent classes for cases, we extend the traditional latent class approach to avoid the LI assumption. The proposed model is a natural extension of pLCM \citep{wu2015} and can be used to test its LI assumption.
                  
                  We assume each child's measurements comprise an observation from a mixture model with component classes that represent the $L$ different pathogens that can cause her pneumonia. One primary goal of analysis is to estimate the probability distribution for these classes. To allow for LD, we introduce \textit{latent} \textit{subclasses} nested within each of the $L+1$ ($L$ case, $1$ control) disease classes. Measurements within a subclass are assumed independent. We refer to the model as a ``nested partially-latent class model" or npLCM and use a Bayesian penalty to encourage small but variable numbers of subclasses that parsimoniously approximate the multivariate discrete dependence and avoid overfitting (Section \ref{sec:prior}).
                  
                  We show that the proposed model is partially-identifiable \citep{gustafson2015bayesian} and incorporate prior knowledge about measurement sensitivities to facilitate Bayesian estimation of the etiologic fractions. The npLCM model is estimated via Markov chain Monte Carlo (MCMC) with designed precision to approximate the posterior distributions of the population etiologic fractions, individual latent state, as well as functions of them, such as the fraction of pneumonia cases caused by bacteria.
                  
                  In Section \ref{sec:formulation}, we formulate our model and discuss its statistical properties. Section \ref{sec:computation} provides details on the posterior sampling algorithm to draw inference based on our model. Section \ref{sec:simulations} illustrates through asymptotic evaluations and finite-sample simulations the benefits of the new model relative to a version that ignores LD.  Section \ref{sec:application} applies the proposed method to PERCH study data. Section \ref{sec:nplcm.discussion} concludes with remarks on the method's advantages, limitations, and future extensions.
                  
                  \section{Nested Partially-Latent Class Model}
                  \label{sec:formulation}
                  In this section, we specify the nested partially-latent class model (npLCM)  and consider its statistical properties using the PERCH study example to make the ideas concrete. Let $\bm{M}_i=(M_{i1},...,M_{iJ})'$ comprise a $J$-dimensional multivariate binary measurement collected for subjects $i=1,...,n_1+n_0$, where the first $n_1$ subjects are cases and the remaining $n_0$ are controls. Let $Y_i=1$ denote a case and $Y_i=0$ denote a control. 
                  
                  \subsection{Measurement Likelihood}
\begin{figure}[H]
                                                                                                                                                                                                                                                                                                                                                                                                                                                                                                                                                                                                                                                                                                                                                                                                                                                                                                                                                                                                                                                                              \begin{center}
                                                                                                                                                                                                                                                                                                                                                                                                                                                                                                                                                                                                                                                                                                                                                                                                                                                                                                                                                                                                                                                                              \includegraphics[width=\textwidth]{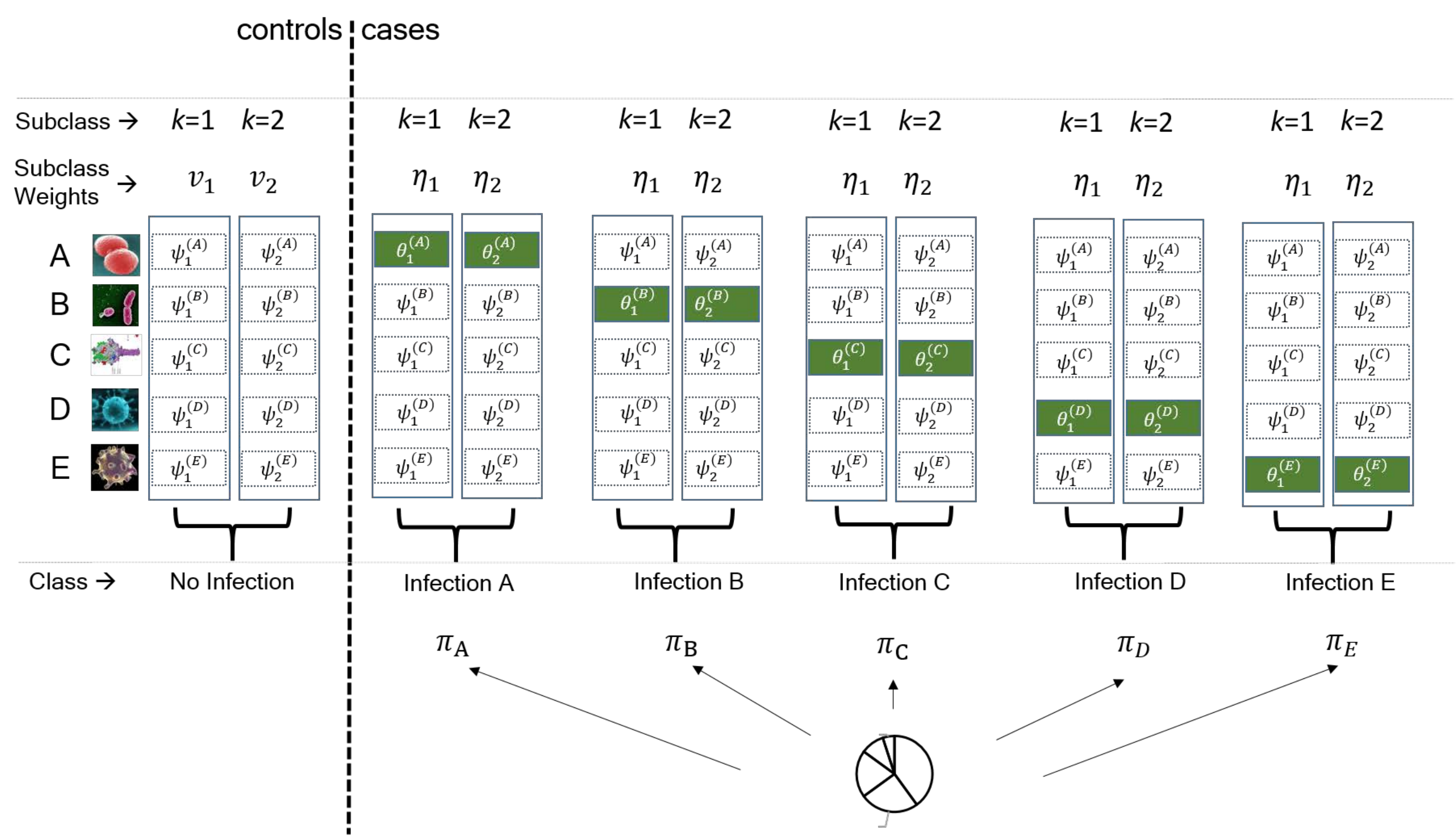}
                                                                                                                                                                                                                                                                                                                                                                                                                                                                                                                                                                                                                                                                                                                                                                                                                                                                                                                                                                                                                                                                              \end{center}
                                                                                                                                                                                                                                                                                                                                                                                                                                                                                                                                                                                                                                                                                                                                                                                                                                                                                                                                                                                                                                                                              \caption{Borrowing measurement characteristics from controls to cases using $K=2$ subclasses for each disease class. Five pathogens (A to E) are measured in this example. }
                                                                                                                                                                                                                                                                                                                                                                                                                                                                                                                                                                                                                                                                                                                                                                                                                                                                                                                                                                                                                                                                              \label{fig:mixture_structure}
                                                                                                                                                                                                                                                                                                                                                                                                                                                                                                                                                                                                                                                                                                                                                                                                                                                                                                                                                                                                                                                                              \end{figure}

                  Figure \ref{fig:mixture_structure} pictures the general structure of the npLCM with $J=5$ measurements, one pathogen per row in the matrix. With $5$ pathogens, there are $6$ classes: one for the control state (pathogen-free) on the left of the dashed vertical line; and $L=5$ case states, one for each possible cause on the right. In the figure, the control measurements have joint distribution that is approximated by a mixture of $K=2$ subclasses, with $K$-dimensional mixing weights $\bnu =(\nu_1, ..., \nu_K)'$. Here $\bpsi_{k} = \{\psi_{k}^{(j)}\}_{1\leq j \leq J}$ is the column vector of false positive rates for measurements $j=1,...,J$, for subclass $k=1,...,K$. The mixing weights of the $K$ subclasses in the case population (right of dashed line) are assumed to be $\bEta=(\eta_1, ..., \eta_K)'$. The {\it etiologic fractions} are the mixing weights for the $L(=J)$ classes in the case population, denoted $\bpi=(\pi_1,...,\pi_L)'$ with $\sum_{\ell=1}^L\pi_{\ell} = 1$. 
                  
                  Throughout the paper, we rely on the scientific assumption that each child's pneumonia is caused by a single primary pathogen. The more general case where disease can be attributed to multiple pathogens is a natural extension (Section \ref{sec:nplcm.discussion}).

                  \subsection{Control Likelihood}
                  The control measurement distribution is assumed to take the form in \cite{Goodman1974}. Mutual dependence is induced by the existence of multiple subclasses, with each subclass having possibly distinct positive rate profiles. Given an unobserved subclass, measurements are assumed to be mutually independent. Marginalizing over the latent subclasses produces dependence for pathogens with different rates across subclasses. The formulation is natural for PERCH given the heterogeneity in the health status of controls.
                  
                  For control $i$, we introduce subclass indicator $Z_i$ that takes value in  $\{1,..., K\}$ and let
                  \begin{alignat}{3}
                  & {\sf sample~ subclass~indicator:}   \quad && Z_i \sim  {\sf Multinomial}(\{1,...,K\}, \bm{\nu})          \quad &&\label{eq:control.subclass.mix}\\
                  & {\sf generate~measurements:}   \quad && M_{ij}\mid Z_i=k \sim  {\sf Bernoulli}(\psi_k^{(j)}), \text{~independently~for~}j=1,...,J, &&\label{eq:control.subclass.mix2}
                  \end{alignat}
                  where $\nu_k = \PP(Z_i = k \mid Y_i=0)$ and $\psi^{(j)}_k = \PP(M_{ij} = 1 \mid Z_i= k, Y_i = 0)$. Here $\bnu$ comprises of the probabilities of a control falling in the subclasses; $\psi_{k}^{(j)}$ is the probability of a positive response within subclass $k$ viewed as an event of false detection for controls and hence is termed the false positive rate (FPR); the FPRs for subclass $k$ are collected in the FPR profile vector $\bpsi_k$ which is then combined by column into the matrix $\bPsi = [\bpsi_1|...|\bpsi_K]$ for all subclasses. The control distribution of the $2^J$ measurement patterns ($\forall \bm{m}\in\{0,1\}^J$) are then given by 
                  \begin{eqnarray}
                  \bm{P}^{0}(\bm{m}) & = & \PP(\bM_i =\bm{m} \mid \bnu, \bPsi, Y_i=0)=\sum_{k=1}^{K} \nu_k \prod_{j=1}^J\left\{\psi_{k}^{(j)}\right\}^{m_{j}}\left\{1-\psi_{k}^{(j)}\right\}^{1-m_{j}}.\label{eq:ctrl_lkd}
                  \end{eqnarray}
                  
                  \subsection{Case Likelihood}
                  \label{sec:case.lkd}
                  For a case with known cause, her vector of binary measurements is again assumed to be generated from a latent $K$-subclass model as for the controls. In PERCH context, motivated by the observation that cases and controls have similar correlation patterns for many pathogen pairs (e.g., Appendix Figure 2), we let the cases share controls' measurement characteristics. To be more precise, given a case's disease class $I_i = \ell_0\in\{1,\ldots, L\}$, with $L=J$, she falls into subclass $k$ with probability $\eta_k$, for $k=1,...,K$. Then subclass $k$'s response probabilities are assumed equal to $\psi^{(j)}_k$ as in controls for $j\neq \ell_0$, and equal to a new parameter $\theta^{(j)}_{k}$ for $j=\ell_0$. That is, an infection by pathogen $j$ may alter the response probabilities in the $j$-th dimension but not others. Since the disease for cases $i$ is in fact unknown, her measurement distribution is a mixture across all $L$ states given by 
                  $\bP^1(\bm{m})=\PP(\bm{M}_{i}=\bm{m}\mid \bm{\pi}, \bm{\eta}, \bm{\Theta},\bm{\Psi}, Y_i=1)$, $\forall {\bm{m}\in \{0,1\}^J}$,
                  \begin{eqnarray}
                  \bm{P}^{1}(\bm{m})& = & \sum_{\ell=1}^L\pi_{\ell}
                  \sum_{k=1}^{K} \left[\eta_k \left\{\theta_{k}^{(\ell)}\right\}^{m_{\ell}}
                                        \left\{1-\theta_{k}^{(\ell}\right\}^{1-m_{\ell}}
                                        \prod_{j\neq \ell}\left\{\psi_{k}^{(j)}\right\}^{m_{j}}
                                        \left\{1-\psi_{k}^{(j)}\right\}^{1-m_{j}}\right],\label{eq:case_lkd}
                  \end{eqnarray}
                  where $\bm{\Theta}$ is a parameter matrix with $(j,k)$-th element $\theta_k^{(j)}$. 
                  
                  We can reformulate (\ref{eq:case_lkd}) by a three-stage generative process similar to (\ref{eq:control.subclass.mix}-\ref{eq:control.subclass.mix2}) by indicators of case disease classes $I_{i}$ and the nested subclasses $Z_{i}$:
                  \begin{alignat}{3}
                  &{\sf sample~class~indicator: }\quad && I_{i}\mid Y_{i}=1 \sim {\sf Multinomial}(\{1,...,L\},\bm{\pi}),&&\label{eq:eti_class}\\
                  &{\sf sample~subclass~indicator:}   \quad && Z_{i}\mid I_{i}=\ell \sim {\sf Multinomial}(\{1,...,K\},\bm{\eta}), \ell=1,...,L &&\label{eq:case.subclass.mix}\\
                  & {\sf generate~measurements: } \quad && M_{ij}\mid Z_{i}=k, I_{i} \overset{}{\sim} {\sf Bernoulli}\left(\theta_k^{(j)}\mathbf{1}_{\{I_{i}=j\}}+\psi_k^{(j)}\mathbf{1}_{\{I_{i}\neq j\}}\right) , &&\label{eq:alter}
                  \end{alignat}
                  independently for $ 1\leq j \leq J$. At the first stage, the vector $\bm{\pi}$ comprises probabilities of a case in class $1$ to $L$ and is the primary target of inference in this paper. Then, the cases' subclass mixing weights $\bEta=(\eta_1, \ldots, \eta_K)'$ determines the probability of a case falling into each subclass. The final stage generates the measurement at the $j$-th dimension: positive with probability $\theta_k^{(j)}$ or $\psi_k^{(j)}$ according as the realized values of $I_{i}$ and $Z_{i}$ in previous steps. Because $\theta_k^{(j)}$  is the probability of true detection for infections caused by pathogen $j$, we term it true positive rate (TPR) and collect them in $\btheta_k=(\theta_{k}^{(1)},\ldots, \theta_{k}^{(J)})'$ for subclass $k$. 
                  
                  Importantly, case and controls' subclass mixing weights ($\bEta$ and $\bnu$) need not be identical. This admits a measurement dependence structure for cases different from that in controls, such as pathogen increased or reduced interactions due to the former's lung infection. We refer to the special case of $\bm{\eta}=\bm{\nu}$ (element-wise equality) as \textit{non-interference submodels}, under which controls and cases of class $j$ have identical distributions of the leave-one-dimension-out measurement vector $\bM_{i[-j]}$.  Further setting $\eta_1=\nu_1=1$, or $K=1$, gives the pLCM.
                  
                  We have assumed cases' latent states categories take value from a complete list of $J$ measured pathogens (i.e., $L=J$). The case likelihood (\ref{eq:case_lkd}) can be extended to account for \textit{other} causes by adding an extra term:
                    \(
                      \pi_{J+1}\sum_{k=1}^{K} \eta_k \left(\prod_{j=1}^J\{\psi_{k}^{(j)}\}^{m_{j}}
                                                            \{1-\psi_{k}^{(j)}\}^{1-m_{j}}\right),
                      \)
                  where $\pi_{J+1} = \PP(I_{i}=J+1)$ is the total etiology fraction of other causes. For a clinically-confirmed pneumonia case, the negative responses on $J$ pathogens by highly-sensitive assays indicate the possibility of other etiologic pathogens.

                  Combining (\ref{eq:ctrl_lkd}) and (\ref{eq:case_lkd}), the joint likelihood across independent subjects is given by
                  \begin{eqnarray}
                  \mathcal{L}(\bm{\pi},\bm{\Theta}, \bm{\Psi}, \bnu,\bm{\eta}; \mathcal{D}) & = &
                    \prod_{i:Y_i=0}\bm{P}^{0}(\bm{M}_i)
                  \prod_{i:Y_{i'}=1}\bm{P}^{1}(\bm{M}_{i'}),\label{eq:total.lkd}
                                          \end{eqnarray}
                                          where $\mathcal{D}$ collects all the measurement data.
                                          
                                          \subsection{Properties}
                                          The proposed model extends pLCM in \cite{wu2015} by adding $(3J+1)(K-1)$ additional parameters compared to the original formulation with the total number of parameters linear in $J$ when $K\ll J$ providing a parsimonious approximation to the case and control joint distributions that require $2(2^J-1)$ parameters in a saturated model. We further reduce the effective number of parameters using a penalty prior (Section \ref{sec:prior}).
                                          
                                          We assumed that the LD of measurements within each case class can be explained by allowing the same number of LI subclasses as in the controls, so that the case subclass measurement parameters can be partly informed by their control counterparts  (see (\ref{eq:alter})). Additional case subclasses can be included once $I_i$ is directly observed for some cases. 
                                          
                                          
                                          In Appendix A, we provide expressions of the marginal means and pairwise associations for multivariate binary measurements given the npLCM likelihood. These formulas are used to study the magnitude of dependence given true parameters and to generate marginal posterior distributions for observables used in model checking, as illustrated in Section \ref{sec:asymp_bias} and \ref{sec:application}.
                                          
                                                                                                                                                                                                                                                                                                                                                                                                                                                                                          
                                                                                                                                                                                                                                                                                                                                                                                                                                                                                          \subsection{Prior Specifications}
                                                                                                                                                                                                                                                                                                                                                                                                                                                                                          \label{sec:prior}
                                                                                                                                                                                                                                                                                                                                                                                                                                                                                          For the npLCM, we specify the prior distributions on unknown parameters as follows:
                                                                                                                                                                                                                                                                                                                                                                                                                                                                                          \begin{eqnarray}
                                                                                                                                                                                                                                                                                                                                                                                                                                                                                          \bm{\pi} & \sim & {\sf Dirichlet}(a_1,\dots,a_{L}),\\
                                                                                                                                                                                                                                                                                                                                                                                                                                                                                          \psi_k^{(j)} &\sim& {\sf Beta}(b_{1kj},b_{2kj}), j = 1, ..., J; k = 1, ..., \infty,\\
                                                                                                                                                                                                                                                                                                                                                                                                                                                                                          \theta_k^{(j)} & \sim & {\sf Beta}(c_{1kj},c_{2kj}), j=1,...,J; k = 1,..., \infty,\\
                                                                                                                                                                                                                                                                                                                                                                                                                                                                                          Z_{i} \mid Y_{i} = 1 &\sim& \sum_{k=1}^{\infty}U_k\prod_{s<k}\left[1-U_{s}\right]\delta_k, ~~~~U_k\sim {\sf Beta}(1,\alpha_1), i=1,...,n_1,\label{eq:case_sb}\\%
                                                                                                                                                                                                                                                                                                                                                                                                                                                                                          Z_i \mid Y_i = 0&\sim & \sum_{k=1}^{\infty}V_k\prod_{s<k}[1-V_{s}]\delta_k, ~~~~V_k\sim {\sf Beta}(1,\alpha_0), i=n_1+1,...,n_1+n_0,\label{eq:ctrl_sb}\\
                                                                                                                                                                                                                                                                                                                                                                                                                                                                                          \alpha_0, \alpha_1 &\sim&{\sf Gamma}(0.25,0.25),\label{eq:stick.break.hyperprior}
                                                                                                                                                                                                                                                                                                                                                                                                                                                                                          \end{eqnarray}
                                                                                                                                                                                                                                                                                                                                                                                                                                                                                          where $\delta_k$ is a point mass on $k$, and prior independence is also assumed among these parameters. As discussed in more detail by \cite{wu2015}, the npLCM is partially identified \citep{Jones2010}. Specifically, the TPRs $\bTheta$ are not fully identified by the model likelihood (\ref{eq:total.lkd}). Therefore, we choose $(c_{1kj},c_{2kj}), \forall k,j$, so that the $2.5\%$ and $97.5\%$ quantiles of the Beta distribution with parameters $(c_{1kj},c_{2kj})$ match the prior minimum and maximum TPR values elicited from pneumonia experts (Section \ref{sec:application}). Otherwise, we use the default value of $1$s for the Beta hyperparameters. Hyperparameters for the etiology prior, $(a_1,...,a_J)'$, are usually $1$s to denote equal and non-informative prior weights for each pathogen if expert prior knowledge is unavailable. 
                                                                                                                                                                                                                                                                                                                                                                                                                                                                                          
                                                                                                                                                                                                                                                                                                                                                                                                                                                                                          Because our goal is to estimate the etiology fractions, $\bm{\pi}$, after marginalizing over subclass indicators ($Z_{i}$), the parameters for the dependence structure within each disease class are nuisance parameters. Therefore, rather than fixing $K$, we let $K$ be a random positive integer and perform model averaging using a prior that encourages small values of $K$ to incorporate its uncertainty into the inference about $\bm{\pi}$ in a parsimonious way.  This prevents model overfitting in finite samples when the observed contingency table for the multivariate binary PERCH measurements has mostly empty cells. In (\ref{eq:case_sb}) and (\ref{eq:ctrl_sb}), we have actually specified stick-breaking priors for both $\bm{\eta}=\left\{U_k\prod_{s<k}\left[1-U_{s}\right]\right\}_{k=1,2,...}$ and $\bnu=\left\{V_k\prod_{s<k}[1-V_{s}]\right\}_{k=1,2,...}$ that on average place decreasing weights on the $k$th subclass as $k$ increases \citep{sethuraman1994}. Appendix B further discusses the use of stick-breaking prior in our model. The priors above are conjugate to the likelihood of unknown parameters, making the Gibbs sampler in Section \ref{sec:computation} conveniently constructed.

                                                                                                                                                                                                                                                                                                                                                                                                                                                                                          \section{Posterior Computations}
                                                                                                                                                                                                                                                                                                                                                                                                                                                                                          \label{sec:computation}
                                                                                                                                                                                                                                                                                                                                                                                                                                                                                          The posterior distributions of the population etiology fraction vector ($\bm{\pi}$), TPRs ($\bTheta$) and FPRs ($\bPsi$)  can be estimated by simulating approximating samples from the joint posterior via Markov chain Monte Carlo (MCMC) algorithms \citep{brooks2011handbook}. Appendix Figure 1 presents the directed acyclic graph (DAG) for the model structure with observed and latent variables in the npLCM. For posterior computation involving stick-breaking priors, without truncation on the number of stick segments, \citet{Walker2007} and \citet{papaspiliopoulos2008retrospective} proposed the slice sampler and retrospective MCMC, respectively. In the following, we develop a simple and efficient blocked Gibbs sampler relying on truncation approximation to the stick-breaking prior distribution \citep[e.g.,][]{ishwaran2001gibbs, gelfand2002computational}. We also include in the sampling algorithms two sets of auxiliary variables, the partially-latent individual class indicator ($I_i$) the nested subclass indicator ($Z_i$). Appendix C shows the algorithm step-by-step.
                                                                                                                                                                                                                                                                                                                                                                                                                                                                                          
                                                                                                                                                                                                                                                                                                                                                                                                                                                                                          All model estimations are performed by the \verb"R" package ``\verb"baker"" (\href{https://github.com/zhenkewu/baker}{https://github.com/zhenkewu/baker}) that interfaces with freely available software \verb"JAGS 3.4.0" \href{http://mcmc-jags.sourceforge.net/}{(http://mcmc-jags.sourceforge.net/)}. Convergence was monitored via MCMC chain histories, auto-correlations, kernel density plots, and Brooks-Gelman-Rubin statistics \citep{brooks1998general}. The statistical results below are based on $10,000$ iterations of burn-in followed by $50,000$ production samples from each of three parallel chains. Samples from every $50$ iterations are retained for inference.
                                                                                                                                                                                                                                                                                                                                                                                                                                                                                          
                                                                                                                                                                                                                                                                                                                                                                                                                                                                                          \section{Asymptotic and Simulation Studies of Nested Partially-Latent Class Models}
                                                                                                                                                                                                                                                                                                                                                                                                                                                                                          \label{sec:simulations}
                                                                                                                                                                                                                                                                                                                                                                                                                                                                                          
                                                                                                                                                                                                                                                                                                                                                                                                                                                                                          This section presents asymptotic and simulation studies to show that for cases like PERCH 1) when the LI assumption is incorrect, a working LI model will estimate $\bpi$ with asymptotic bias; 2) fitting the LD model to data generated with LI does not lose too much efficiency using sparse priors  on subclass indicators; and 3) compared to the LI model, the LD model produces $95\%$ credible intervals for $\bpi$ with better actual coverage rates.
                                                                                                                                                                                                                                                                                                                                                                                                                                                                                          
                                                                                                                                                                                                                                                                                                                                                                                                                                                                                          \subsection{Asymptotic Bias Evaluations}
                                                                                                                                                                                                                                                                                                                                                                                                                                                                                          \label{sec:asymp_bias}
                                                                                                                                                                                                                                                                                                                                                                                                                                                                                          
                                                                                                                                                                                                                                                                                                                                                                                                                                                                                          We first evaluate the asymptotic bias of using a working LI model (pLCM) in the estimation of $\bpi$ for $J$ causes, i.e., $L=J$. Under the LI assumption, let the maximum likelihood estimator be $\widehat{\bpi}_N = (\widehat{\pi}_{N,1}, \ldots, \widehat{\pi}_{N,J-1}, \widehat{\pi}_{N,J})'$, where the $J$-th etiologic fraction is estimated as $\widehat{\pi}_{N,J} = 1-\sum_{\ell\neq J}\widehat{\pi}_{N,\ell}$ and  $N=n_1+n_0$ is the total sample size. The estimator $\{\widehat{\bpi}_{N,\ell}\}_{1\leq \ell\leq J-1}$ will converge to the first $L-1$ components of the parameter vector $\bomega^*=(\bpi^*_1, \ldots, \bpi^*_{J-1}, \bpsi^*{\marg})'$ that minimizes the Kullback-Leibler information criterion, or equivalently, to the $\bomega^*$ that satisfies
                                                                                                                                                                                                                                                                                                                                                                                                                                                                                          \begin{equation}
                                                                                                                                                                                                                                                                                                                                                                                                                                                                                          \EE_{\bOmega_0}\left\{\frac{\partial}{\partial \bomega}\log \P_{\bomega}(\bM_i\mid Y_i)\Bigr|_{\bomega^*}\right\}=0,\label{eq:ee}
                                                                                                                                                                                                                                                                                                                                                                                                                                                                                          \end{equation}
                                                                                                                                                                                                                                                                                                                                                                                                                                                                                          where $\bOmega_o = (\bpi_o, \bPsi_o, \bTheta_o)$ is the collection of etiologic fractions, FPRs and TPRs in the true data generating mechanism, and $\PP_{\bomega}(\bM_i\mid Y_i)$ is the likelihood of the pLCM given by
                                                                                                                                                                                                                                                                                                                                                                                                                                                                                          \begin{dmath}
                                                                                                                                                                                                                                                                                                                                                                                                                                                                                          \sum_{\ell\neq J}\pi_\ell \cdot f_\ell(\bM_i, \bpsi\marg) + \left(1-\sum_{\ell\neq J}\pi_{\ell}\right)\cdot f_J(\bM_i, \bpsi\marg),
                                                                                                                                                                                                                                                                                                                                                                                                                                                                                          \end{dmath}
                                                                                                                                                                                                                                                                                                                                                                                                                                                                                          where \(f_\ell(\bm{m}, \bpsi\marg) = (\theta_\ell\marg)^{\bm{m}_{j}}(1-\theta_\ell\marg)^{1-\bm{m}_{j}}\prod_{j\neq \ell}(\psi_{j}\marg)^{\bm{m}_{j}}(1-\psi_{j}\marg)^{1-\bm{m}_{j}}\) for cases $Y_i=1$; and $\prod_{j=1}^J(\psi_{j}\marg)^{\bm{m}_{ij}}(1-\psi_{j}\marg)^{1-\bm{m}_{ij}}$ for controls $Y_i=0$. We also fix at the true values the \textit{marginal} sensitivities $\theta_j\marg = \sum_{k=1}^K\theta^{(j)}_{o,k}\eta_k, j=1,\ldots, J$, to eliminate the partial-identifiability issue and to focus on asymptotic bias evaluations. Our calculation of the expectation in (\ref{eq:ee}) assumed equal case and control sample sizes, and could be easily modified for other sampling ratios. Further, \cite{white1982maximum} also established the asymptotic normality of the estimator:
                                                                                                                                                                                                                                                                                                                                                                                                                                                                                          \(\sqrt{N}(\widehat{\bomega}_N-\bomega^*)\overset{d}{\rightarrow} \mathcal{N}\left(\bm{0}, A(\bomega^*)^{-1}B(\bomega^*)A(\bomega^*)^{-1}\right),\)
                                                                                                                                                                                                                                                                                                                                                                                                                                                                                          where 
                                                                                                                                                                                                                                                                                                                                                                                                                                                                                          \begin{equation}
                                                                                                                                                                                                                                                                                                                                                                                                                                                                                          A(\bomega^*) = - \EE_{\bOmega_0}\left\{\frac{\partial^2}{\partial \bomega^2}\log \P_{\bomega}(\bM_i\mid Y_i)\Bigr|_{\bomega^*}\right\}, 
                                                                                                                                                                                                                                                                                                                                                                                                                                                                                          B(\bomega^*) = \VV_{\bOmega_0}\left\{\frac{\partial}{\partial \bomega}\log \P_{\bomega}(\bM_i\mid Y_i)\Bigr|_{\bomega^*}\right\}.\label{eq:robust.var}
                                                                                                                                                                                                                                                                                                                                                                                                                                                                                          \end{equation}
                                                                                                                                                                                                                                                                                                                                                                                                                                                                                          The robust variance of $\widehat{\bomega}_N$ is calculated as $V^*_R =  N^{-1}A^{-1}(\bm{\omega}^*)B(\bm{\omega}^*)A^{-1}(\bm{\omega}^*)$ by approximating the variance operator in $B(\bomega^*)$ using empirical samples. We use Monte Carlo method with $10^7$ samples from $[\bM_i\mid Y_i = y]$, for $y=0, 1$, to evaluate the expectation in (\ref{eq:ee}) and then numerically solve it for its root to obtain $\bomega^{*}$. We then calculate (\ref{eq:robust.var}) by plugging in $\bomega^{*}$ and evaluating the expectation using $10^7$ Monte Carlo samples.
                                                                                                                                                                                                                                                                                                                                                                                                                                                                                          
                                                                                                                                                                                                                                                                                                                                                                                                                                                                                          The strength of LD given disease class determines the estimation bias. When the true data generating mechanism is close to independence, the working LI model estimates of $\bpi$ are close to being asymptotically unbiased. To illustrate, we quantify the asymptotic bias for $J=5$ binary measures (pathogens A, B, C, D and E). We generate Monte Carlo samples from the true data generating mechanisms with varying degrees of LD, while fixing the etiologic fraction $\bpi_o= (0.5, 0.2, 0.15, 0.1, 0.05)'$ to mimic what is seen in PERCH. We create associations among measurements by defining two subclasses ($K=2$) for each of the $6$ disease states (controls plus 5 disease classes for cases). We consider two scenarios of measurement parameters $(\bPsi, \bTheta)$: ({\sf I}) little LD --- small between-subclass differences in positive rates; ({\sf II}) substantial LD --- large differences (see Appendix F). 
                                                                                                                                                                                                                                                                                                                                                                                                                                                                                          
\begin{figure}[!p]
                  \begin{center}
                  \includegraphics[width=\linewidth]{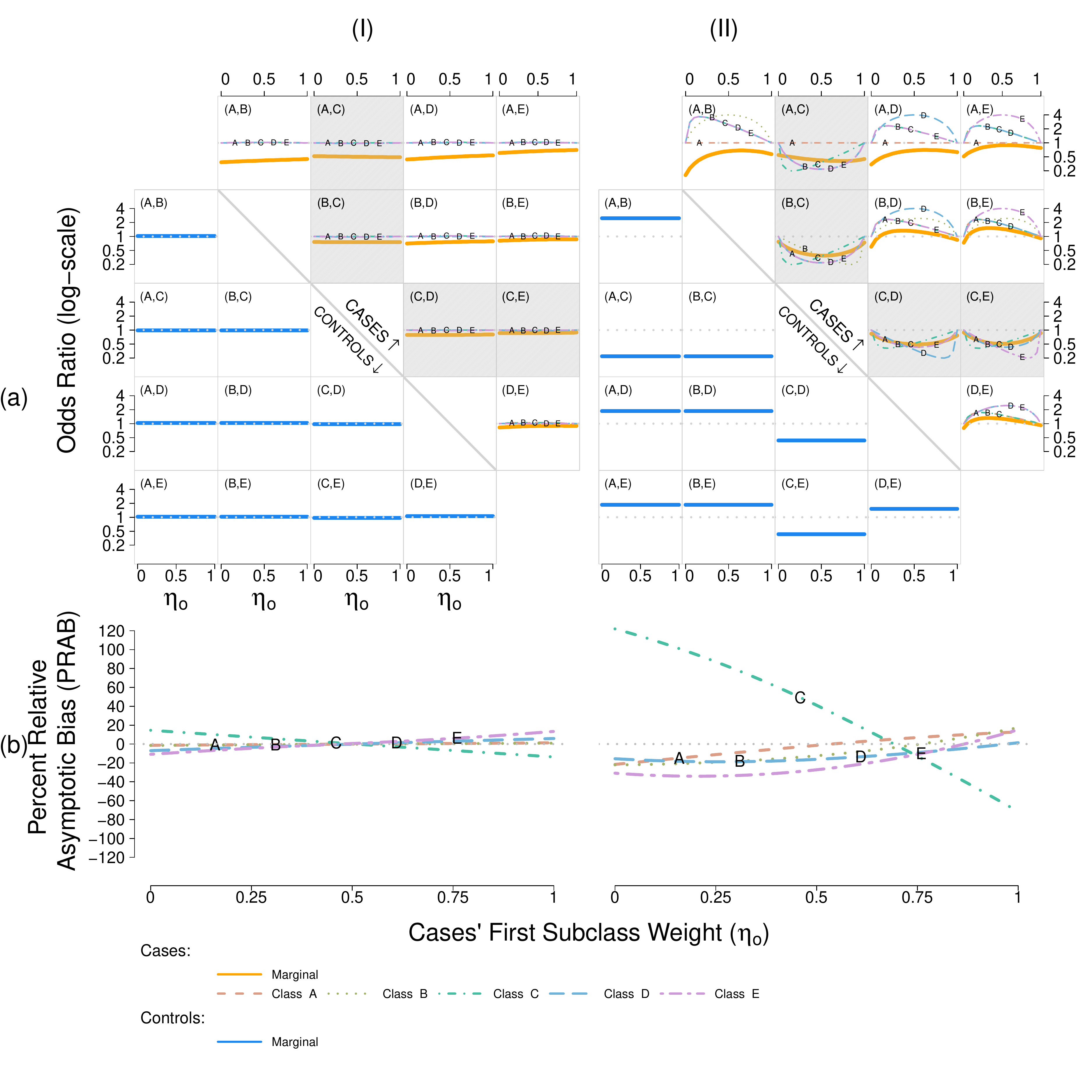}
                  \end{center}
                  \caption[]{In Scenario {\sf \textbf{I}}-{\sf \textbf{II}}, \\
                    \textit{Top (a)}: The true data generating mechanism summarized by pairwise odds ratios for cases (upper right, solid lines) and controls (lower left, solid lines) as the cases' first subclass weight ($\eta_0$) increases from 0 to 1. The pairwise odds ratios \textit{within} each case class are shown by non-solid lines (legend at bottom). Pairwise independence is represented by the dotted horizontal lines for reference. The correlations of C with others are highlighted in shaded cells.\\
                    \textit{Bottom (b)}: Percent relative asymptotic bias (PRAB) for estimating etiology fractions using working local independence (LI) model when the truth varies across a range of local dependence (LD) settings parametrized by $\eta_o$.}
                    \label{fig:asymp_bias_with_condOR}
                    \end{figure}

                                                                                                                                                                                                                                                                                                                                                                                                                                                                                          The subclass weights characterize the degree of LD. We assume controls and cases fall into the first subclass with probability $\nu_o=0.5$ and $\eta_o\in[0,1]$, respectively. Row (a) of Figure \ref{fig:asymp_bias_with_condOR} summarizes both the marginal and within-class dependence for Scenario {\sf I} and {\sf II}. The marginal associations are stronger in Scenario {\sf II} (solid curves). Note that the within-class odds ratio curves leave and return to 1 and remain above or below 1 as $\eta_o$ increases from 0 to 1 (non-solid lines labelled by A-E in small panels), because when all the weight is on one of the two subclasses, the true data generating
                                                                                                                                                                                                                                                                                                                                                                                                                                                                                          mechanism satisfies LI. In particular, the equality $\eta_o=\nu_o(=0.5)$ represents identical LD structures (non-interference submodels) for cases and controls, with deviations from it indicating differential dependence patterns.

                                                                                                                                                                                                                                                                                                                                                                                                                                                                                          Row (b) of Figure \ref{fig:asymp_bias_with_condOR} summarizes the results by the percent relative asymptotic bias  (PRAB) at all $\eta_o$ values for each etiologic fraction, $(\pi^*_\ell-\pi_{o,\ell})/\pi_{o,\ell}\times100\%$. The working LI model produces PRABs less than $13\%$ in magnitude in Scenario {\sf I}. Given small asymptotic biases, we also obtain good estimates of precision produced by the working LI model with the ratios for model-based variance $V^*_M = N^{-1}A^{-1}(\bomega^*)$ versus the robust variance $V^*_R$ between $0.97^2$ and $1.05^2$ for A-E. The two variances are mathematically identical at arbitrary parameter values if marginal FPRs ($\bpsi\marg$) are known.
                                                                                                                                                                                                                                                                                                                                                                                                                                                                                          
                                                                                                                                                                                                                                                                                                                                                                                                                                                                                          The asymptotic bias is large under strong LD as in Scenario {\sf II}. For example, the working LI model overestimates $\pi_{oC}$ with $121.3\%$ relative bias at $\eta_o=0$ for its failure to account for the strong LD among controls. When the case LD is more similar to controls at $\eta_o=0.5$, the PRAB  is $40.5\%$. This is because the measurement on C is negatively associated with the measurements on B, D, or E given disease class B, D, or E, i.e. mutual inhibition (see shaded cells in Figure \ref{fig:asymp_bias_with_condOR}, a-{\sf II}), leading to the case pattern $\bM_{i} = (1,0,1,1,0)'$ observed twice as frequently as expected by a working LI model. When they are further assigned with the highest likelihood to cause C under the working LI model, the upward bias results. 
                                                                                                                                                                                                                                                                                                                                                                                                                                                                                          
                                                                                                                                                                                                                                                                                                                                                                                                                                                                                          \subsection{Bayesian Fitting in Finite Samples}
                                                                                                                                                                                                                                                                                                                                                                                                                                                                                          In finite samples, one can fit the larger LD model that \textit{a priori} encourages a small number of subclasses. Extra subclasses can be used if the measurements have rich multivariate associations. Through simulations, we compare Bayes estimates of etiologic fractions obtained from the npLCM and pLCM.  We generate $T = 1,000$ datasets with sample size $n_1=n_0=500$ under Scenario {\sf I} and {\sf II}. We fit the npLCM (truncation level $K^*=5$ subclasses) and pLCM ($K=1$) to each data set using informative Beta priors on the true positive rates ($\{\theta^{(j)}_k\}$) with $0.5$ and $0.99$ as the $2.5\%$ and $97.5\%$ quantiles mimicing PERCH study.
                                                                                                                                                                                                                                                                                                                                                                                                                                                                                          
                                                                                                                                                                                                                                                                                                                                                                                                                                                                                          We view the Bayes estimates as functionals of data and assess their frequentist properties, such as bias and variance \citep[e.g.][]{efron2015frequentist}. We define the repeated-sampling bias of the posterior mean and its mean squared error (MSE) respectively as \(
                                                                                                                                                                                                                                                                                                                                                                                                                                                                                          \lim_T T^{-1}\sum_{t=1}^T \left\{\mean{\pi}_\ell^{(t)}-\pi_{o,\ell}\right\}\), and \(\lim_T T^{-1}\sum_{t=1}^T (\mean{\pi}_\ell^{(t)}-\pi_{o,\ell})^2, \ell=A, \ldots, E,\)
                                                                                                                                                                                                                                                                                                                                                                                                                                                                                          where $\mean{\pi}_\ell^{(t)}=\EE\{\pi_\ell\mid \cD^{(t)},\cM\}$ is the posterior mean taken with respect to the posterior distribution of $\bpi$ given the $t$-th simulated data set $\cD^{(t)}$ and model $\cM$.
                                                                                                                                                                                                                                                                                                                                                                                                                                                                                          
                                                                                                                                                                                                                                                                                                                                                                                                                                                                                          The top panel of Table \ref{table:pie_estimation_comp} compares the estimation bias by posterior means obtained from the two models. For a data set with finite sample size, estimation bias can arise from random sampling, model mis-specification or the prior, for which the first is averaged out by replication. The non-zero biases seen here reflect likelihood mis-specification and the influence of the prior. When the likelihood is correctly specified, only biases from priors remain. In Scenario {\sf II} with strong LD, the npLCM performs much better. For example, the LI assumption (pLCM) results in an upward bias of $26.2\%$ for C at $\eta_o=0$, as well as other highlighted biases greater than $10\%$ in magnitude. In Scenario {\sf I} with weak LD, the biases from both models are negligible ($-1.9\%$ $\sim$ $1.9\%$).
                                                                                                                                                                                                                                                                                                                                                                                                                                                                                          
                                                                                                                                                                                                                                                                                                                                                                                                                                                                                          When the truth is close to LI, the npLCM is comprablely efficient to pLCM for almost all settings. The bottom panel of Table \ref{table:pie_estimation_comp} shows the the ratio of MSEs for pLCM versus npLCM. In Scenario {\sf I}, the ratios are close to $1$ indicating the npLCM has efficiently used stick-breaking to strike the balance between estimation bias and variance. In Scenario {\sf II}, compared to the pLCM, the npLCM produced smaller MSEs for C at all $\eta_o$ values,  where the advantage is largely explained by smaller biases.
                                                                                                                                                                                                                                                                                                                                                                                                                                                                                          
                                                                                                                                                                                                                                                                                                                                                                                                                                                                                          The npLCM also produces $95\%$ credible intervals (CI) with near-nominal empirical coverage rates. For example, Appendix Table 1 highlights that the substantial under-coverages ($<80\%$) only occurred when assuming LI. Because of the extra variability from the informative priors on the TPRs, the CIs are conservative in Scenario {\sf I} for both models. The over-coverage of both models is largely due to the assumed variances in the TPR parameters.
                                                                                                                                                                                                                                                                                                                                                                                                                                                                                          
                                                                                                                                                                                                                                                                                                                                                                                                                                                                                          \section{Analysis of PERCH Data}
                                                                                                                                                                                                                                                                                                                                                                                                                                                                                          \label{sec:application}
                                                                                                                                                                                                                                                                                                                                                                                                                                                                                          The Pneumonia Etiology Research for Child Health (PERCH) study is a standardized
                                                                                                                                                                                                                                                                                                                                                                                                                                                                                          and comprehensive case-control study that has enrolled over $4,000$ patients hospitalized for severe or very severe pneumonia and over $5,000$ controls selected randomly from the community, frequency-matched on age in each month. Its objective is to evaluate etiologic agents causing severe and very severe
                                                                                                                                                                                                                                                                                                                                                                                                                                                                                          pneumonia among hospitalized children aged $1$-$59$ months in seven low and middle income countries with a significant burden of childhood pneumonia and a range of epidemiologic characteristics \citep{Levine2012}. More details about the PERCH design can be found in \cite{Deloria2012}. 
                                                                                                                                                                                                                                                                                                                                                                                                                                                                                          
                                                                                                                                                                                                                                                                                                                                                                                                                                                                                          Using preliminary PERCH data from one site, we focus on PCR assays on nasopharyngeal (NP) specimens for cases and controls. We illustrate the advantage of the npLCM in accounting for measurement LD, with improved efficiency, better empirical fit, and more valid etiology estimation. Results for all seven countries will be reported elsewhere upon study completion. Included in the current illustrative analysis are NPPCR data for $592$ cases and $613$ controls on $6$ species of pathogens (abbreviations and full names in Appendix F). 
                                                                                                                                                                                                                                                                                                                                                                                                                                                                                          
                                                                                                                                                                                                                                                                                                                                                                                                                                                                                          We have compared the population etiology fractions, ${\bpi}$, estimated separately by two methods: the pLCM and the npLCM with subclass truncation level $K^*=10$. The npLCM results are similar when larger values of $K^*$s are used.  As discussed in Section \ref{sec:prior}, we need expert prior knowledge on the sensitivities for posterior inference by both methods; we used elicited sensitivity priors from laboratory experts with range $50\sim 99.5\%$. Given our focus on $6$ leading pathogens, we include the ``other" cause for completeness as discussed in Section \ref{sec:case.lkd}. 
                                                                                                                                                                                                                                                                                                                                                                                                                                                                                          
                                                                                                                                                                                                                                                                                                                                                                                                                                                                                          Strong LD is present in the analyzed data, with statistically significant log odds ratios observed for $6$ out of $30$ pathogen pairs among cases and controls, ranging from $-2.47~(\text{s.e.:~} 1.01)$ to $1.67~(\text{s.e.:~} 0.39)$, and also by noting that under LI assumption we expect $0.05\times 30=1.5(\pm2.4)$ such pairs. In addition, as noted in \cite{berger1987testing} and \cite{dunson2009nonparametric}, the interval null hypothesis $H_0: \max_k \eta_k >1-\epsilon$, is useful for detecting deviations from the point null of exact LI.  We choose $\epsilon = 0.05$ based on experience in simulation studies and to permit deviations from LI so small as to be non-significant in our application. The largest subclass weight is estimated with $95\%$ CI $(0.65, 0.89)$ for the cases and $(0.27, 0.75)$ for the controls, again suggesting non-negligible LD in the data.
                                                                                                                                                                                                                                                                                                                                                                                                                                                                                          
                                                                                                                                                                                                                                                                                                                                                                                                                                                                                          Figure 4(a) compares the results obtained from the pLCM (left boxes) and npLCM (right boxes). Each vertical box-and-whisker shows the marginal posterior mean (solid dot) and median (segment within box),  with $95\%$ credible interval (CI; between whisker endpoints) and $50\%$ CI (between top and bottom box edges) of the etiologic fraction for each pathogen listed on the horizontal bar. The two approaches produce differences in the posterior means of etiologic fractions between $-9.9\%$ and  $9.5\%$. Half of the largest increase in \pname{RHINO}, from $5.2~(\text{95}\%\text{~CI:~} 0.3 \sim 17.9)\%$ to $15.1~(5.9 \sim 27.5)\%$ is explained by its increase in predicted individual etiologies for cases with the NPPCR data {\sf 000010} (Figure \ref{fig:app.ind.pred}, bottom left).
                                                                                                                                                                                                                                                                                                                                                                                                                                                                                                                                                                                                                                                                                                                                                                                                                                                                                                                                                                                                                                                                              
                                                                                                                                                                                                                                                                                                                                                                                                                                                                                                                                                                                                                                                                                                                                                                                                                                                                                                                                                                                                                                                                              The npLCM also provides a better empirical fit. We have compared the posterior predictive distributions \citep{gelman1996posterior} of the frequencies of common NP measurement patterns to the observed values separately in the cases and the controls. Among cases (left panel in Figure 4(b)), for example, the npLCM adequately predicts the observed frequencies of the 2nd and 6th most common case patterns ({\sf 000001}: $12.5\%$; {\sf 000100}: $5.4\%$) by accounting for the negative associations of RSV with other pathogens with the log odds ratios ranging from $-3.37$ to $-0.12$ ($3$ out of $5$ statistically significant). 
                                                                                                                                                                                                                                                                                                                                                                                                                                                                                                                                                                                                                                                                                                                                                                                                                                                                                                                                                                                                                                                                              
                                                                                                                                                                                                                                                                                                                                                                                                                                                                                                                                                                                                                                                                                                                                                                                                                                                                                                                                                                                                                                                                              We also examine the pairwise associations by calculating the standardized LOR difference (SLORD) defined to be the observed LOR for a pair of measurements minus the mean LOR for the predictive distribution value from each method divided by the standard deviation of the LOR predictive distribution. Appendix Figure 3 shows $9$ pairs of pathogens that have statistically significant deviations of model predicted LORs from the observed ones for the pLCM and only $3$ pairs for the npLCM. A blank cell indicates a good model prediction for the observed pairwise LOR ($\mid SLORD \mid <2$). The npLCM achieves a better fit by noting that, for a well-fitting model, we expect $1.5 (\pm2.4)$ non-blank cells. The associations between pairs of measurements ({\small \sf HMPV-A/B},{\small \sf RSV}) and ({\small \sf PARA-1},{\small \sf RSV}) are not expected in either model, although npLCM does better. In the PERCH study, we observed that seasonal variation in the rate of detection for {\small \sf RSV}, {\small \sf HMPV-A/B} and {\small \sf PARA-1} were out of phase and seasonal regression adjustment, discussed elsewhere, can sensibly account for this negative association.
                                                                                                                                                                                                                                                                                                                                                                                                                                                                                                                                                                                                                                                                                                                                                                                                                                                                                                                                                                                                                                                                              
                                                                                                                                                                                                                                                                                                                                                                                                                                                                                                                                                                                                                                                                                                                                                                                                                                                                                                                                                                                                                                                                              \section{Discussion}
                                                                                                                                                                                                                                                                                                                                                                                                                                                                                                                                                                                                                                                                                                                                                                                                                                                                                                                                                                                                                                                                              \label{sec:nplcm.discussion}
                                                                                                                                                                                                                                                                                                                                                                                                                                                                                                                                                                                                                                                                                                                                                                                                                                                                                                                                                                                                                                                                              In this paper, we derived and tested a nested pLCM to allow for local dependence among binary observations given class membership. We compare this new model with a special case that depends on local independence in terms of asymptotic and finite sample size properties. The npLCM reduces large-sample estimation bias, retains the estimation efficiency and gives more valid inferences about $\bpi$ than the pLCM. The npLCM family also makes it possible to study the sensitivity of scientific findings to the LI  assumption when pLCM is used.
                                                                                                                                                                                                                                                                                                                                                                                                                                                                                                                                                                                                                                                                                                                                                                                                                                                                                                                                                                                                                                                                              
                                                                                                                                                                                                                                                                                                                                                                                                                                                                                                                                                                                                                                                                                                                                                                                                                                                                                                                                                                                                                                                                              The model first approximates the probability distribution for the control measurements by a mixture of product Bernoulli distributions with mixing weights penalized towards a mixture with few components. The estimated control dependence structure is then applied to the case model with modifications that represent the influence of the latent disease state. This valuable information from controls may help distinguish competing models for the local dependence among measurements and warrants further studies \citep[e.g.][]{albert2001latent}. 
                                                                                                                                                                                                                                                                                                                                                                                                                                                                                                                                                                                                                                                                                                                                                                                                                                                                                                                                                                                                                                                                              
                                                                                                                                                                                                                                                                                                                                                                                                                                                                                                                                                                                                                                                                                                                                                                                                                                                                                                                                                                                                                                                                              
                                                                                                                                                                                                                                                                                                                                                                                                                                                                                                                                                                                                                                                                                                                                                                                                                                                                                                                                                                                                                                                                              
                                                                                                                                                                                                                                                                                                                                                                                                                                                                                                                                                                                                                                                                                                                                                                                                                                                                                                                                                                                                                                                                              In the analysis of $6$ leading pathogens from the PERCH study, {\small \sf RSV} is estimated to be the most prevalent infectious cause of childhood pneumonia except the ``other" category. That evidence is robust to the LD assumption. Accounting for LD structure leads to notable increases in etiologic fraction estimates of two pathogens and decrease in another. The npLCM can also integrate extra measurements of better qualities, for example, blood culture tests for bacteria that have near-perfect specificities to inform TPRs and improve efficiency \citep{Hammitt2012b}.
                                                                                                                                                                                                                                                                                                                                                                                                                                                                                                                                                                                                                                                                                                                                                                                                                                                                                                                                                                                                                                                                              
                                                                                                                                                                                                                                                                                                                                                                                                                                                                                                                                                                                                                                                                                                                                                                                                                                                                                                                                                                                                                                                                              In this paper, we assumed a single primary cause for each pneumonia case in the npLCM. This framework can be extended from a single to multiple causes by using a latent vector for case $i$, $\bm{I}_i\in{\{0,1\}}^J$, where $I_{ij}=1$ indicates pathogen $j$ is a component cause. For example, \cite{Hoff2005} used Dirichlet process mixture models to identify multiple abnormal genomic locations that are jointly responsible for each case's disease, but using case-only data with LI assumption. Alternatively, one can place an exponential penalty on the number of causes \citep[e.g.,][]{Zhang2007}, or use conditionally specified models $[I_{ij} = 1\mid \bm{I}_{i[-j]}, \bm{X}_{ij}]$ to characterize the interactions among pathogens \citep{besag1974spatial}, where $\bm{X}_{ij}$ is a vector of covariates predictive for pathogen $j$ being a cause in case $i$. The computational cost to fit these models increases substantially because the search space for the latent vector $\bm{I}_i$ expands exponentially in $J$. Development of efficient and reliable posterior sampling algorithms can allow investigators to assess the evidence of multiple-pathogen etiologies as more measurements accrue.  
                                                                                                                                                                                                                                                                                                                                                                                                                                                                                                                                                                                                                                                                                                                                                                                                                                                                                                                                                                                                                                                                              
                                                                                                                                                                                                                                                                                                                                                                                                                                                                                                                                                                                                                                                                                                                                                                                                                                                                                                                                                                                                                                                                              A second extension of the npLCM family motivated by PERCH is to allow the etiology distribution and false positive rates to depend upon covariates. For example, season, child's age and HIV status. Regression versions for npLCM have been implemented and are the subject of current study. 
                                                                                                                                                                                                                                                                                                                                                                                                                                                                                                                                                                                                                                                                                                                                                                                                                                                                                                                                                                                                                                                                              
                                                                                                                                                                                                                                                                                                                                                                                                                                                                                                                                                                                                                                                                                                                                                                                                                                                                                                                                                                                                                                                                              Finally, \cite{wu2015} derived the pLCM model to be used with a combination of direct measurements of cases' lungs without error and peripheral measures of cases and controls with error. With gold-standard data, this analyses is an example of supervised learning. The npLCM can be used in the same way. In the PERCH application, we rely entirely on peripheral samples, so the analyses is largely unsupervised. Robustness of inferences to model assumptions is critical.

                                                                                                                                                                                                                                                                                                                                                                                                                                                                                                                                                                                                                                                                                                                                                                                                                                                                                                                                                                                                                                                                              \section*{Supplementary Material}

                                                                                                                                                                                                                                                                                                                                                                                                                                                                                                                                                                                                                                                                                                                                                                                                                                                                                                                                                                                                                                                                              The reader is referred to the Supplementary Material at the end of this paper for technical appendices and additional simulations referenced in Sections 2, 3, 4 and 5.
                                                                                                                                                                                                                                                                                                                                                                                                                                                                                                                                                                                                                                                                                                                                                                                                                                                                                                                                                                                                                                                                              \section*{Acknowledgments}
                                                                                                                                                                                                                                                                                                                                                                                                                                                                                                                                                                                                                                                                                                                                                                                                                                                                                                                                                                                                                                                                              We thank the members of the larger PERCH Study Group for discussions that helped shape the statistical approach presented herein, and the study participants. We also thank the members of PERCH Expert Group who provided external advice.

                                                                                                                                                                                                                                                                                                                                                                                                                                                                                                                                                                                                                                                                                                                                                                                                                                                                                                                                                                                                                                                                              \bibliographystyle{biorefs}
                                                                                                                                                                                                                                                                                                                                                                                                                                                                                                                                                                                                                                                                                                                                                                                                                                                                                                                                                                                                                                                                              \bibliography{nplcm}

                                                                                                                                                                                                                                                                                                                                                                                                                                                                                                                                                                                                                                                                                                                                                                                                                                                                                                                                                                                                                                                                              \newpage
                                                                                                                                                                                                                                                                                                                                                                                                                                                                                                                                                                                                                                                                                                                                                                                                                                                                                                                                                                                                                                                                              

                  \FloatBarrier
                  \newpage
                  
                  
                  \newpage

                  \newpage 
                  \begin{table}[!p]
                  \caption[]{Comparison of Bayes estimates of etiology fractions obtained from npLCM ({\ sf np}) and pLCM ({\sf p}). \textit{Top}:  direct bias of the posterior mean ($\mean{\pi}_\ell-\pi_{o,\ell}$); \textit{Bottom}: ratio of mean squared errors (MSE) for pLCM vs npLCM.  All numbers are averaged across $1,000$ replications and multiplied by $100$.}
                  \label{table:pie_estimation_comp}
                  \centering
                  \renewcommand{\arraystretch}{1.1}
                  \sf
                  \begin{tabular}{cccccccccccc}
                  \hline
                  && & \multicolumn{9}{c}{Truth: Cases' First Subclass Weight ($\eta_o$)}\\
                  \vspace{0.1em}
                  &	  & Model & 0 & & 0.25 &  & 0.5 & & 0.75 & &  1  \\ 
                  \hline
                  &\underline{Class} & &\multicolumn{9}{c}{\multirow{1}{*}{\underline{100$\times$\msd{Bias}{ Standard Error}}} }\\
                  \multirow{10}{*}{\textbf{I}} &	\multirow{2}{*}{$A$} &  np 
                  & \msd{ -0.8}{  0.1}  && \msd{ -0.5}{  0.1}  && \msd{ -0.2}{  0.1}  && \msd{  0.1}{  0.1}  && \msd{  0.4}{  0.1} \\ 
                  &&  p 
                  & \msd{ -1.1}{  0.1}  && \msd{ -0.7}{  0.1}  && \msd{ -0.3}{  0.1}  && \msd{ -0.1}{  0.1}  && \msd{  0.0}{  0.1} \\ 
                  & \multirow{2}{*}{$B$} &  np 
                  & \msd{ -0.6}{  0.1}  && \msd{ -0.5}{  0.1}  && \msd{ -0.4}{  0.1}  && \msd{ -0.5}{  0.1}  && \msd{ -0.4}{  0.1} \\ 
                  &&  p 
                  & \msd{ -0.6}{  0.1}  && \msd{ -0.5}{  0.1}  && \msd{ -0.6}{  0.1}  && \msd{ -0.5}{  0.1}  && \msd{ -0.3}{  0.1} \\ 
                  & \multirow{2}{*}{$C$} &  np 
                  & \msd{  1.4}{  0.1}  && \msd{  0.7}{  0.1}  && \msd{ -0.1}{  0.1}  && \msd{ -0.9}{  0.1}  && \msd{ -1.7}{  0.1} \\ 
                  &&  p 
                  & \msd{  1.9}{  0.1}  && \msd{  0.8}{  0.1}  && \msd{ -0.1}{  0.1}  && \msd{ -0.9}{  0.1}  && \msd{ -1.9}{  0.1} \\ 
                  & \multirow{2}{*}{$D$} &  np 
                  & \msd{ -0.1}{  0.1}  && \msd{  0.1}{  0.1}  && \msd{  0.4}{  0.1}  && \msd{  0.6}{  0.1}  && \msd{  0.9}{  0.1} \\ 
                  &&  p 
                  & \msd{ -0.2}{  0.1}  && \msd{  0.3}{  0.1}  && \msd{  0.5}{  0.1}  && \msd{  0.7}{  0.1}  && \msd{  1.1}{  0.1} \\ 
                  & \multirow{2}{*}{$E$} &  np 
                  & \msd{  0.0}{  0.1}  && \msd{  0.2}{  0.1}  && \msd{  0.3}{  0.1}  && \msd{  0.6}{  0.1}  && \msd{  0.7}{  0.1} \\ 
                  &&  p 
                  & \msd{  0.0}{  0.0}  && \msd{  0.2}{  0.1}  && \msd{  0.5}{  0.1}  && \msd{  0.8}{  0.1}  && \msd{  1.0}{  0.1} \\ \cline{2-12} \ 
                  \multirow{10}{*}{\textbf{II}} &	\multirow{2}{*}{$A$} &  np 
                  & \msd{  4.5}{  0.1}  && \msd{  5.7}{  0.1}  && \msd{  5.5}{  0.1}  && \msd{  3.5}{  0.1}  && \msd{  0.5}{  0.1} \\ 
                  &&  p 
                  & \msd{ -3.6}{  0.1}  && \msd{  0.2}{  0.1}  && \msd{  3.0}{  0.1}  && \msd{  5.0}{  0.1}  && \msd{  5.5}{  0.1} \\ 
                  & \multirow{2}{*}{$B$} &  np 
                  & \msd{ -5.7}{  0.1}  && \msd{ -6.1}{  0.1}  && \msd{ -4.9}{  0.1}  && \msd{ -2.1}{  0.1}  && \msd{  1.9}{  0.1} \\ 
                  &&  p 
                  & \msd{\textbf{-13.5}}{  0.1}  && \msd{ -8.5}{  0.1}  && \msd{ -4.3}{  0.1}  && \msd{ -0.3}{  0.1}  && \msd{  4.1}{  0.1} \\ 
                  & \multirow{2}{*}{$C$} &  np 
                  & \msd{  4.5}{  0.1}  && \msd{  4.1}{  0.1}  && \msd{  2.1}{  0.1}  && \msd{ -1.0}{  0.1}  && \msd{ -6.2}{  0.1} \\ 
                  &&  p 
                  & \msd{ \textbf{26.2}}{  0.1}  && \msd{ \textbf{13.6}}{  0.1}  && \msd{  3.7}{  0.1}  && \msd{ -4.8}{  0.1}  && \msd{\textbf{-12.5}}{  0.0} \\ 
                  & \multirow{2}{*}{$D$} &  np 
                  & \msd{ -2.4}{  0.1}  && \msd{ -2.5}{  0.1}  && \msd{ -1.7}{  0.1}  && \msd{ -0.4}{  0.1}  && \msd{  2.1}{  0.1} \\ 
                  &&  p 
                  & \msd{ -5.8}{  0.0}  && \msd{ -3.3}{  0.1}  && \msd{ -1.6}{  0.1}  && \msd{ -0.2}{  0.1}  && \msd{  1.3}{  0.1} \\ 
                  & \multirow{2}{*}{$E$} &  np 
                  & \msd{ -1.0}{  0.0}  && \msd{ -1.3}{  0.0}  && \msd{ -1.0}{  0.0}  && \msd{ -0.1}{  0.1}  && \msd{  1.6}{  0.1} \\ 
                  &&  p 
                  & \msd{ -3.2}{  0.0}  && \msd{ -1.9}{  0.0}  && \msd{ -0.8}{  0.1}  && \msd{  0.4}{  0.1}  && \msd{  1.7}{  0.1} \\ 
                  
                  \hline
                  \hline
                  & \underline{Class}& &\multicolumn{9}{c}{\multirow{1}{*}{\underline{\msd{100$\times$Ratio of MSE}{ Standard Error}} }}\\
                  
                  \multirow{5}{*}{\textbf{I}} &	\multirow{1}{*}{$A$} &   
                  & \msd{  94}{   6}  && \msd{ 115}{   7}  && \msd{ 100}{   6}  && \msd{  92}{   6}  && \msd{  91}{   6} \\ 
                  & \multirow{1}{*}{$B$} &  
                  & \msd{ 105}{   6}  && \msd{  94}{   6}  && \msd{  98}{   6}  && \msd{  96}{   6}  && \msd{  91}{   6} \\ 
                  & \multirow{1}{*}{$C$} &  
                  & \msd{ 114}{   7}  && \msd{ 101}{   6}  && \msd{  93}{   5}  && \msd{  93}{   5}  && \msd{  90}{   5} \\ 
                  & \multirow{1}{*}{$D$} &  
                  & \msd{ 104}{   6}  && \msd{ 105}{   6}  && \msd{  96}{   6}  && \msd{  97}{   6}  && \msd{ 108}{   7} \\ 
                  & \multirow{1}{*}{$E$} &   
                  & \msd{  97}{   4}  && \msd{  96}{   6}  && \msd{ 124}{   7}  && \msd{  98}{   6}  && \msd{ 119}{   7} \\ 
                  \cline{2-12} \\
                  \multirow{5}{*}{\textbf{II}} &	\multirow{1}{*}{$A$} &   
                  & \msd{  82}{   4}  && \msd{  25}{   1}  && \msd{  47}{   2}  && \msd{ 115}{   6}  && \msd{ 221}{  12} \\ 
                  & \multirow{1}{*}{$B$} &  
                  & \msd{ 516}{  11}  && \msd{ 177}{   5}  && \msd{  80}{   3}  && \msd{  62}{   4}  && \msd{ 140}{   8} \\ 
                  & \multirow{1}{*}{$C$} &  
                  & \msd{2379}{  77}  && \msd{ 711}{  26}  && \msd{ 131}{   7}  && \msd{ 268}{  13}  && \msd{ 357}{   8} \\ 
                  & \multirow{1}{*}{$D$} &  
                  & \msd{ 397}{  14}  && \msd{ 152}{   6}  && \msd{  94}{   5}  && \msd{  79}{   4}  && \msd{  60}{   4} \\ 
                  & \multirow{1}{*}{$E$} &   
                  & \msd{ 357}{  13}  && \msd{ 151}{   6}  && \msd{ 102}{   5}  && \msd{  95}{   6}  && \msd{  82}{   5} \\ 
                  
                  \hline\hline
                  
                  \end{tabular}
                  \end{table}
                  
                  
                  \newpage
                  \begin{figure}[!p]
                  \label{fig:app.compare}
                  \centering     
                  \subfigure[]{\label{fig:etiology.estimates_02}\includegraphics[width=\linewidth]{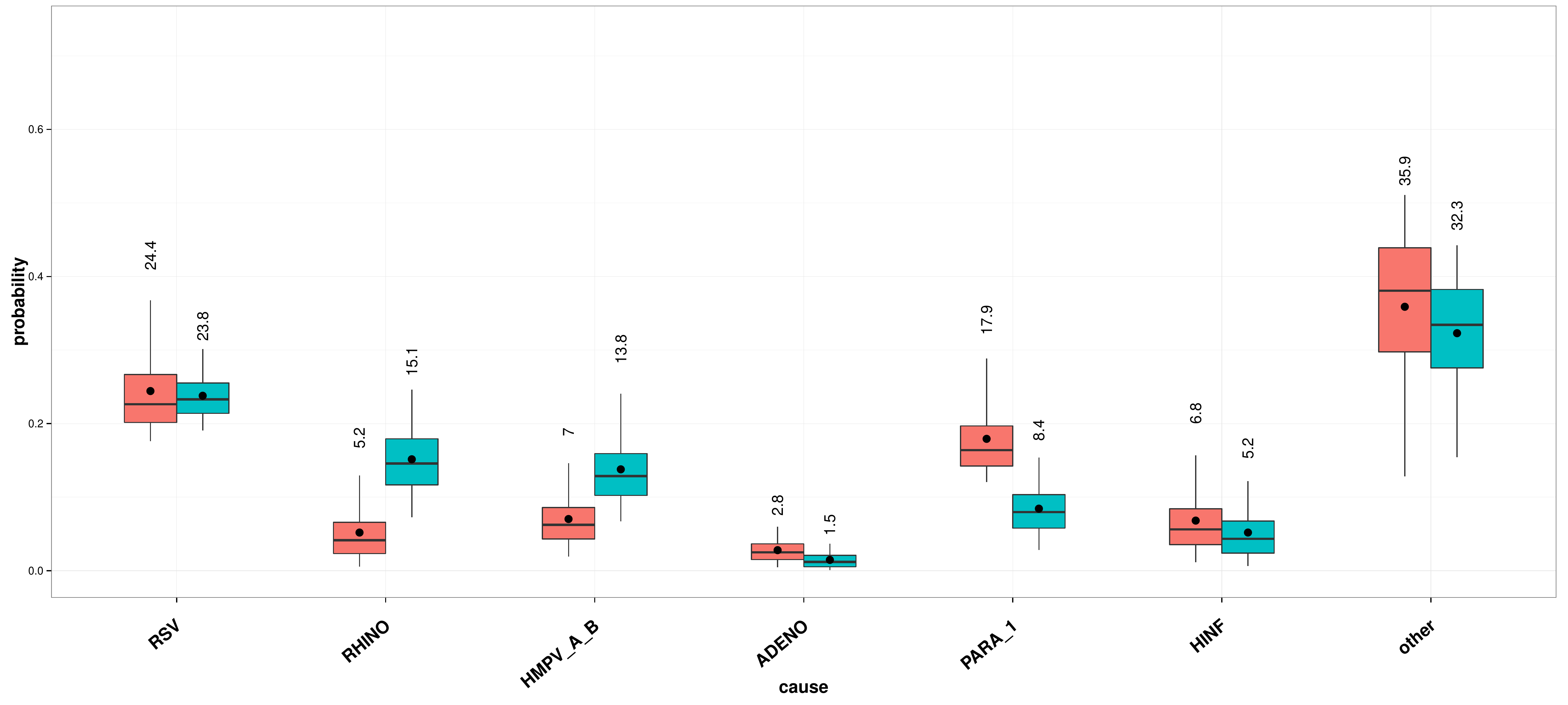}}
                  \subfigure[]{\label{fig:checking.common_pattern_02}\includegraphics[width=\linewidth]{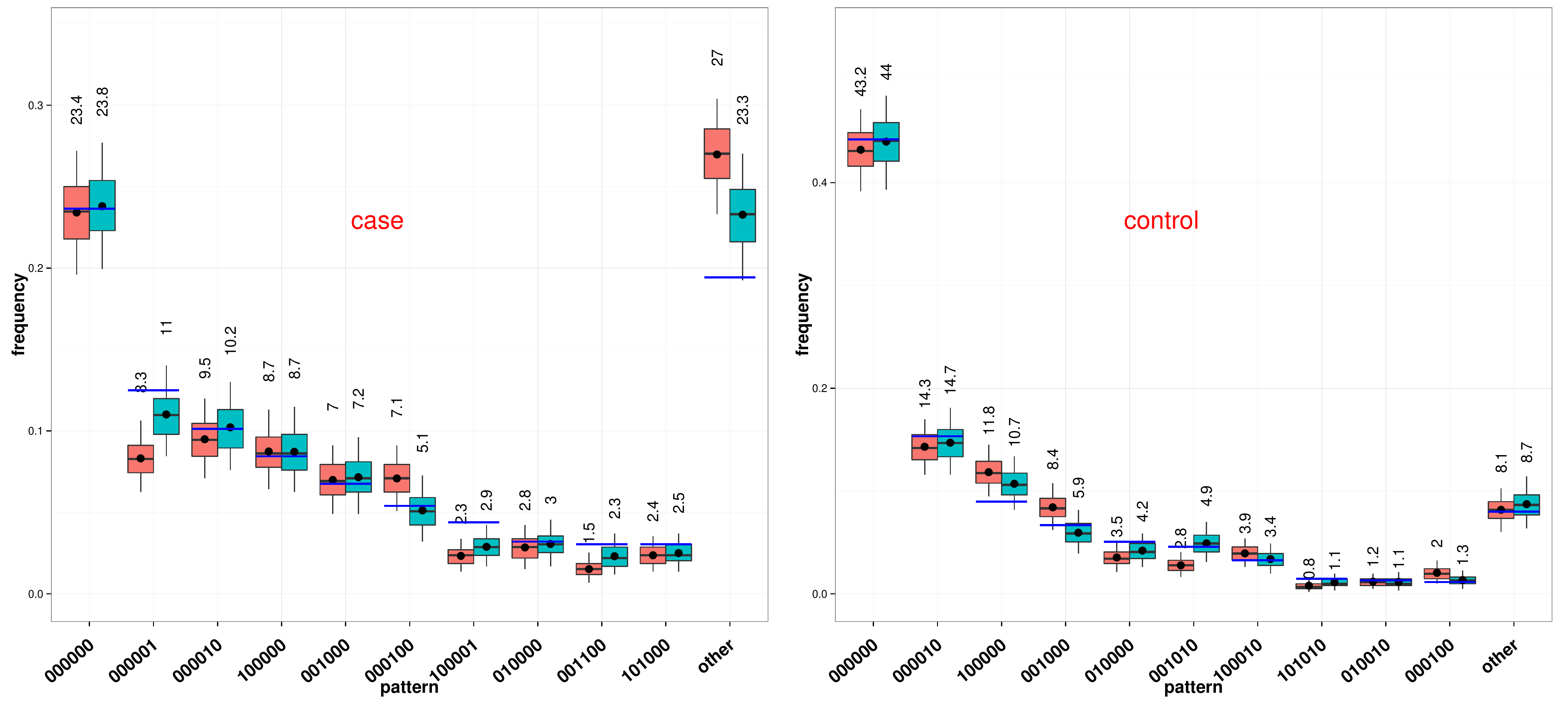}}
                  \caption{\textit{Top}: Comparison of the posterior distributions of $\bpi$ between the pLCM (left) and npLCM (right); The numbers above are the posterior means ($\times 100$). \textit{Bottom}: Posterior predictive distributions (PPD) for 10 most frequent multivariate binary patterns separately for cases (left panel) and controls (right panel). The observed frequencies are overlayed as short segments across pairs of box-and-whiskers; the means of the PPDs ($\times 100$) are shown above them in actual numbers.}
                  \end{figure}
                  
                  \newpage
                  \begin{figure}[!p]
                  \captionsetup[subfigure]{labelformat=empty}
                  \centering     
                  \includegraphics[width=\linewidth]{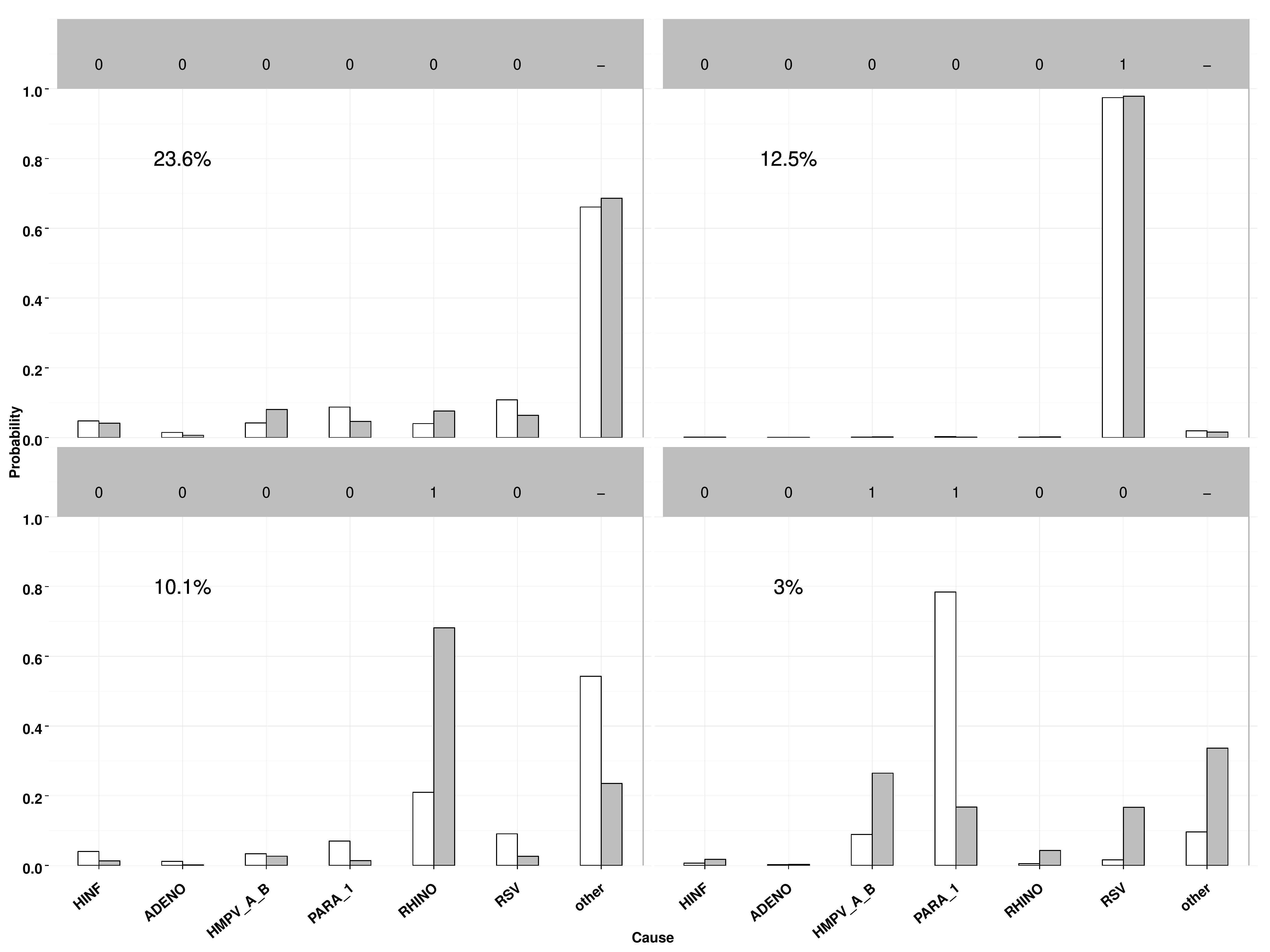}
                  \caption{Individual etiology distribution estimated by the empirical distribution of MCMC samples of the disease class indicator. Here four NPPCR data patterns are represented by the binary codes at the top (no measurements on ``other" causes hence left as ``-"), with its observed frequency marked beneath. The height of a bar represents the probability of a case caused by each of the $7$ causes labelled on the horizontal axis. For each cause, paired bars compare the estimates from the pLCM (left) and the npLCM (right); Extra predictions are in Appendix Figure 4.}
                  \label{fig:app.ind.pred}
                  \end{figure}


                  %
                  %

                  \FloatBarrier
                  \newpage 
                  
                  \appendix
                  
                  \setcounter{equation}{0}
                  \renewcommand{\theequation}{A\arabic{equation}}
                  \setcounter{figure}{0}
                  \renewcommand{\figurename}{}
                  \renewcommand{\tablename}{}
                  \renewcommand{\thefigure}{Appendix Figure \arabic{figure}}
                  \renewcommand{\thetable}{Appendix Table \arabic{table}}
                  \renewcommand{\thesection}{Appendix \Alph{section}}
                  \renewcommand{\thesubsection}{\thesection.\arabic{subsection}}
                  

                  \section{Mean and Covariance Structure}
                  \label{sec:mean_covariance}
                  In this section, we present and discuss formulas for the model-based marginal observation rates and pairwise log odds ratios among cases and controls. They can be readily modified to accommodate ``other" causes as discussed in Section 2.3 of the main text.
                  
                  \subsection{Marginal Observation Rate}
                  The marginal observation rates are given by
                  \begin{eqnarray}
                  \PP(M_{i'j}=1\mid Y_{i'}=1) &=& \pi_j\sum_{k=1}^K\theta_k^{(j)}\eta_k+(1-\pi_j)\left\{\sum_{k=1}^K \psi^{(j)}_k\eta_k\right\},\label{eq:marg.mean1}\\
                  \PP(M_{ij}=1\mid Y_i=0) &=&\sum_{k=1}^K \psi^{(j)}_k\nu_k.\label{eq:marg.mean2}
                  \end{eqnarray}
                  
                  In the context of childhood pneumonia problem, Equation (\ref{eq:marg.mean1}) indicates that the observed rate of pathogen $j$ among cases comprises of two parts: cases whose disease is caused by pathogen $j$ for which the observation is a true positive event, and those whose disease is caused by another pathogen for which the observation is a false positive. 
                  
                  The case and control mean observation rates for pathogen $j$ are equal (i.e., non-interference submodel; Section 2.3 of main text) when either of Condition (I) or (II) below holds.
                  \[(\text{I})~\psi^{(j)}_1  = \cdots = \psi^{(j)}_K=\psi^{(j)} \text{~and~} \sum_{k=1}^K\theta_k^{(j)}\eta_k = \psi^{(j)};\]
                  \[(\text{II})~\bEta  = \bnu, \text{~and~}\sum_{k=1}^K\left[\theta_k^{(j)}\eta_k-\psi^{(j)}_k\nu_k\right]= 0.\]
                  
                  The first part of Condition (I) says that the binary response on dimension $j$ is constant across subclasses among controls, which implies independence of $j$-th dimension's measurement to other dimensions. The second part says, within the $j$th disease class, the marginal observation rate of dimension $j$ equals the control rate.

                  Condition (II) means case and control subclass weights are equal and the observation rate in the $j$th case class equals that in the controls.

                  \subsection{Marginal Pairwise Log Odds Ratios}
                  The marginal pairwise log odds ratio $\omega_{j\ell}$ for pathogen pair ($j,\ell$) among cases is given by:
                  \begin{eqnarray}
                  \omega_{j\ell} & = & \log\left\{\frac{\PP(M_{ij}=1,M_{i\ell}=1)\PP(M_{ij}=0,M_{i\ell}=0)}{\PP(M_{ij}=1,M_{i\ell}=0)\PP(M_{ij}=0,M_{i\ell}=1)}\right\}\nonumber\\
                  &=& \log \left(\sum_{c=1}^{J}\pi_c\left[\sum_{k=1}^K\left\{\theta_k^{(j)}\right\}^{\mathbf{1}\{c=j\}}\left\{\psi_k^{(j)}\right\}^{\mathbf{1}\{c\neq j\}}\left\{\theta_k^{(\ell)}\right\}^{\mathbf{1}\{c=\ell\}}\left\{\psi_k^{(\ell)}\right\}^{\mathbf{1}\{c\neq \ell\}}\eta_k\right]\right)\nonumber\\
                  &&- \log \left(\sum_{c=1}^{J}\pi_c\left[\sum_{k=1}^K\left\{1-\theta_k^{(j)}\right\}^{\mathbf{1}\{c=j\}}\left\{1-\psi_k^{(j)}\right\}^{\mathbf{1}\{c\neq j\}}\left\{\theta_k^{(\ell)}\right\}^{\mathbf{1}\{c=\ell\}}\left\{\psi_k^{(\ell)}\right\}^{\mathbf{1}\{c\neq \ell\}}\eta_k\right]\right)\nonumber\\
                  &&+ \log \left(\sum_{c=1}^{J}\pi_c\left[\sum_{k=1}^K\left\{1-\theta_k^{(j)}\right\}^{\mathbf{1}\{c=j\}}\left\{1-\psi_k^{(j)}\right\}^{\mathbf{1}\{c\neq j\}}\left\{1-\theta_k^{(\ell)}\right\}^{\mathbf{1}\{c=\ell\}}\left\{1-\psi_k^{(\ell)}\right\}^{\mathbf{1}\{c\neq \ell\}}\eta_k\right]\right)\nonumber\\
                  &&- \log \left(\sum_{c=1}^{J}\pi_c\left[\sum_{k=1}^K\left\{\theta_k^{(j)}\right\}^{\mathbf{1}\{c=j\}}\left\{\psi_k^{(j)}\right\}^{\mathbf{1}\{c\neq j\}}\left\{1-\theta_k^{(\ell)}\right\}^{\mathbf{1}\{c=\ell\}}\left\{1-\psi_k^{(\ell)}\right\}^{\mathbf{1}\{c\neq \ell\}}\eta_k\right]\right).\label{eq:pairwise.logOR}
                  \end{eqnarray}
                  
                  Setting $K=1$ in the formula gives log odds ratios for a locally independent model \citep{wu2015}. When $K>1$, suppose nearly all of pneumonia is caused by pathogen $j$: $\pi_j\approx 1$, we calculate $\omega_{j\ell}$ under two scenarios: 
                  \begin{enumerate}
                  \item[a)] If the true positive rates for pathogen $j$ across subclasses, i.e. $\theta_k^{(j)}, k=1,...,K$, are equal, then $\omega_{j\ell}\approx 0$, that is, we have approximate marginal independence between measurements on the $j$th pathogen and the rest among the cases;
                  \item[b)] If the number of subclasses $K=2$ and true positive rates $\theta_k^{(j)}, k=1,2$ are very different, say, 1 versus 0 as an extreme example, we can show that $\omega_{j\ell}=\logit(\psi_1^{(\ell)})-\logit(\psi_2^{(\ell)})$, which means the pairwise log odds ratio between pathogen $j$ and $\ell$ among cases is determined by the variation of control subclass FPRs for the $\ell$th pathogen.
                  \end{enumerate}
                  
                  %
                  
                  \section{Stick-Breaking Prior}
                  \label{sec:stick_breaking}
                  This section briefly discusses the stick-breaking priors used in the Bayesian inference for the nested partially-latent class models. A stick-breaking mixture model in theory has countably infinite number of subclasses. However, because the $\nu_k$ and $\eta_k$ decrease exponentially quickly in $k$, {\it a priori}, we expect that only a small number of subclasses will be used to model the data. The expected number of subclasses from a stick-breaking prior is logarithmic in the number of observations \citep{hjort2010bayesian}. This is different
                  than a finite mixture model, which uses a fixed number of clusters to model the
                  data. In the stick-breaking mixture model, the actual number of clusters used to model data is not fixed, and can be automatically inferred from data using the usual Bayesian posterior inference framework \citep{neal2000markov}. 
                  
                  Equations (2.12)-(2.14) in the text place exchangeable prior weight on the subclasses. Following \cite{ishwaran2002approximate}, in our computations, we truncate the infinite sum to the first $K^*$ terms with $K^*$ sufficiently large to balance computing speed and approximating performance of the model. In our simulations and data application $K^*=10$ is usually deemed adequate. Most subclass measurement profiles are not assigned with meaningful weights either in the simulations or in data application, so that a small number of effective subclasses are usually sufficient for approximation. Also, by placing hyperpriors on stick-breaking parameters $\alpha_0$ and $\alpha_1$ as in Equation (2.14) in the text, we can let the data inform us about the desired sparsity level for approximating the probability contingency tables for the control and each disease class. A small value of the estimate $\widehat{\alpha}_0$ ($\widehat{\alpha}_1$) suggests that only a small number of subclasses are necessary for the controls (cases). We have chosen hyperparameters in the Gamma hyperpriors for $\alpha_1$ and $\alpha_1$ to be $(0.25,0.25)$ which gives good parameter estimation performance in simulations.
                  
                  \section{Gibbs Sampler Algorithm}

                  We propose the following MCMC sampling steps, assuming the truncation level is $K^*=K$:
                  \begin{enumerate}
                  \item Update the class indicator $I_{i'}$ for cases $i'=1,...,n_1$, from a multinomial distribution with probabilities
                  \begin{eqnarray*}
                  \lefteqn{\PP(I_{i'}=j\mid \cdots)=p^{(j)}_{i'} \propto  [\Mb_{i'}\mid Z_{i'},
                    \bTheta,\bPsi, I_{i'}=j][Z_{i'}\mid \bm{\eta}, I_{i'}=j][I_{i'}=j\mid \bpi]}\\
                      &\propto &\left\{\theta_{Z_{i'}}^{(j)}\right\}^{M_{i'j}}\left\{1-\theta_{Z_{i'}}^{(j)}\right\}^{1-M_{i'j}}\prod_{l\neq j} \left\{\psi_{Z_{i'}}^{(l)}\right\}^{M_{i'l}}\left\{1-\psi_{Z_{i'}}^{(l)}\right\}^{1-M_{i'l}}\cdot \eta_{Z_{i'}} \cdot \pi_j,
            \end{eqnarray*}
            for $j=1,...,J$.
            
            \item   Update subclass indicators $Z_{i'}$ for case $i'=1,...,n_1$, from a multinomial distribution with probabilities
            \begin{eqnarray*}
            \PP(Z_{i'}=k \mid \cdots)& = & q_{i'k} \propto  [\Mb_{i'}\mid Z_{i'}, I_{i'}, \bm{\bTheta}, \bm{\Psi}][Z_{i'}\mid I_{i'}, \bEta]\\
              & \propto &\eta_{k}\cdot \left\{\theta^{(I_{i'})}_{k}\right\}^{M_{i'I_{i'}}}\left\{1-\theta^{(I_{i'})}_{k}\right\}^{1-M_{i'I_{i'}}}\prod_{l\neq I_{i'}}\left\{\psi^{(l)}_{k}\right\}^{M_{i'l}}\left\{1-\psi^{(l)}_{k}\right\}^{1-M_{i'l}}.
                                                                                                            \end{eqnarray*}
                                                                                                            
                                                                                                            Update subclass indicators $Z_{i}$ for control $i=n_1+1,...,n_1+n_0$, from a multinomial distribution with probabilities
                                                                                                            \begin{eqnarray*}
                                                                                                            \PP(Z_i=k\mid \cdots) &=& q_{ik}\propto  [\Mb_i\mid Z_i=k, \bPsi][Z_i=k\mid \bm{\nu}]\\
                                                                                                            &\propto &\nu_k\cdot  \prod_{j=1}^J\left\{\psi_k^{(j)}\right\}^{M_{ij}}\left\{1-\psi_k^{(j)}\right\}^{1-M_{ij}}, k=1,...,K.
                                                                                                            \end{eqnarray*}

                                                                                                            \item Update the case subclass weights $\bm{\eta}$ for $j=1,..., J$ from 
                                                                                                            \begin{eqnarray*}
                                                                                                            pr(\bm{\eta}\mid \cdots)  \propto  \prod_{i': I_{i'}=j}[Z_{i'}\mid \bm{\eta}, I_{i'}][\bm{\eta}\mid \alpha_1]\nonumber
                \end{eqnarray*}
                which can be accomplished by first setting $u_K^{*}=1$ and sampling
                \begin{eqnarray*}
                u_k^{*} &\sim &{\sf Beta}\left(1+z'_{k}, \alpha_1+\sum_{l=k+1}^K z'_l\right), k=1,...,K-1,
                \end{eqnarray*}
                where $z'_k$ is the number of cases assigned to subclass $k$ in class $j$. We write 
                \[z'_k=\#\left\{i':Y_{i'}=1, Z_{i'}=k, I_{i'}=j\right\},\] for $k=1,...,K-1$, where ``$\#A$" counts the number of elements in set $A$. We then construct $\eta_1=u_k^{*}$, $\eta_k = u_k^{*}\prod_{l=1}^{k-1}\left\{1-u^{*}_l\right\}$, $k=2,...,K$.
                
                \item Update the control subclass weights $\bm{\nu}=(\nu_1,...,\nu_K)^T$ from 
                \begin{eqnarray*}
                pr(\bm{\nu}\mid \cdots ) 
                &\propto &\prod_{i:Y_i=0}[Z_i\mid \bm{\nu}]\cdot [\bm{\nu}\mid \alpha_0],
                \end{eqnarray*}
                which can be accomplished by first setting $v_K^{*}=1$ and sampling
                \begin{eqnarray*}
                v_k^{*} &\sim &{\sf Beta}\left(1+z_k, \alpha_0+\sum_{l=k+1}^K z_k\right), k=1,...,K-1,
                \end{eqnarray*}
                where $z_k$ is the number of controls assigned to subclass $k$, and then constructing $\nu_1=v_k^{*}$, $\nu_k = v_k^{*}\prod_{l=1}^{k-1}(1-v^{*}_l)$, $k=2,...,K$.
                
                \item Update concentration parameter $\alpha_0$ and $\alpha_1$  for stick-breaking prior from 
                \begin{eqnarray}
                pr(\alpha_0\mid \cdots)\propto  [\bm{\nu}\mid \alpha_0] [\alpha_0]
                &\propto & \alpha_0^{K-1}\exp(-\alpha_0 \cdot r)\cdot pr(\alpha_0),\nonumber
                \end{eqnarray}
                where $r =-\left\{\sum_{k=1}^{K-1}\log(1-\nu^*_k)\right\} $. If conditionally conjugate prior for $\alpha_0$ is used, i.e. $\alpha_0\sim{\sf Gamma}(a_{\alpha_0},b_{\alpha_0})$ with mean $a_{\alpha_0}/b_{\alpha_0}$ and variance $a_{\alpha_0}/b^2_{\alpha_0}$, then the full conditional distribution reduces to \({\sf Gamma}\left(a_{\alpha_0}+K-1, b_{\alpha_0}+r\right).\) Similarly for $\alpha_1$ with $\bnu$ replaced by $\bEta$ and $(a_{\alpha_0}, b_{\alpha_0})$ replaced by $(a_{\alpha_1}, b_{\alpha_1})$.
                
                \item Update the vector of subclass TPR for $j=1,...,J$ from
                \begin{eqnarray*}
                pr(\bm{\theta}^{(j)}\mid \cdots) &\propto & \prod_{\left\{i': I_{i'}=j \right\} }[\Mb_{i'}\mid \bm{\theta}^{(j)}, Z_{i'},I_{i'}][\bm{\theta}^{(j)}]\nonumber\\
                  &\propto &\prod_{k=1}^K\left\{\theta^{(j)}_k\right\}^{m^{(j)}_{k1}}\left\{1-\theta^{(j)}_k\right\}^{m^{(j)}_{k0}}\cdot [\bm{\theta}^{(j)}] ,\label{eq:theta.cond}
                  \end{eqnarray*}
                  where $m^{(j)}_{kc}=\#\{i': Y_{i'}=1, Z_{i'}=k, I_{i'}=j, M_{i'j}=c\}$, $c=0,1$. If prior for TPRs are independent Beta distributions, then this is a product of Beta distributions.
                                                                                                   
                                                                                                   \item Update subclass-specific FPRs $\psi^{(j)}_k$ for $j=1,...,J$, $k=1,...,K$ from
                                                                                                   \begin{eqnarray*}
                                                                                                   pr({\psi}_k^{(j)}\mid \cdots )
                                                                                                   &\propto & \prod_{i': Y_{i'}=1, I_{i'}\neq j, Z_{i'}=k} [M_{i'j}\mid \bm{\psi}^{(j)}, Z_{i'}, I_{i'} ]\prod_{i: Y_i=0}[M_{ij}\mid \bm{\psi}^{(j)}, Z_{i}]\cdot [{\psi}_k^{(j)}]\\
                                                                                                     &\propto &\left\{{\psi}_k^{(j)}\right\}^{s^{(-j)}_{k1}}\left\{1-{\psi}_k^{(j)}\right\}^{s^{(-j)}_{k0}}\cdot pr(\psi_k^{(j)}), 
                                                                                                     \end{eqnarray*}
                                                                                                     where $s^{(-j)}_{kc}=\#\{i': Y_{i'}=1, Z_{i'}=k, I_{i'}\neq j, M_{i'j}=c\}+\#\{i: Y_{i}=0, Z_i=k, M_{ij}=c\}$, for $c=0,1$. If the prior on FPRs are {\sf Beta}$(a_1,b_1)$, then the above conditional distribution is {\sf Beta}$(a_1+s^{(j)}_{k1},b_1+s^{(j)}_{k0}) $.
                                                                                                                                                            \item Update $\bm{\pi}$ from  ${\sf Dirichlet}\left(d_1+t^{(j)},...,d_J+t^{(j)}\right),$
                                                                                                                                                              where $t^{(j)}$ is the number of cases assigned to class $j$, i.e. $t^{(j)}=\#\{i': Y_{i'}=1, I_{i'}=j\}$, $j=1,..,J$.
                                                                                                                                                            \end{enumerate}
                                                                                                                                                            
                                                                                                                                                            \section{Directed Acyclic Graph for Nested Partially-Latent Class Models}
                                                                                                                                                            
                                                                                                                                                            This section illustrates the model structure of nested partially-latent class models using a directed acyclic graph (DAG) and provides some details on posterior inference. 
                                                                                                                                                            
                                                                                                                                                            \label{sec:DAG}
                                                                                                                                                            \begin{figure}[!htp]
                                                                                                                                                            \begin{center}
                                                                                                                                                            \includegraphics[width=\textwidth]{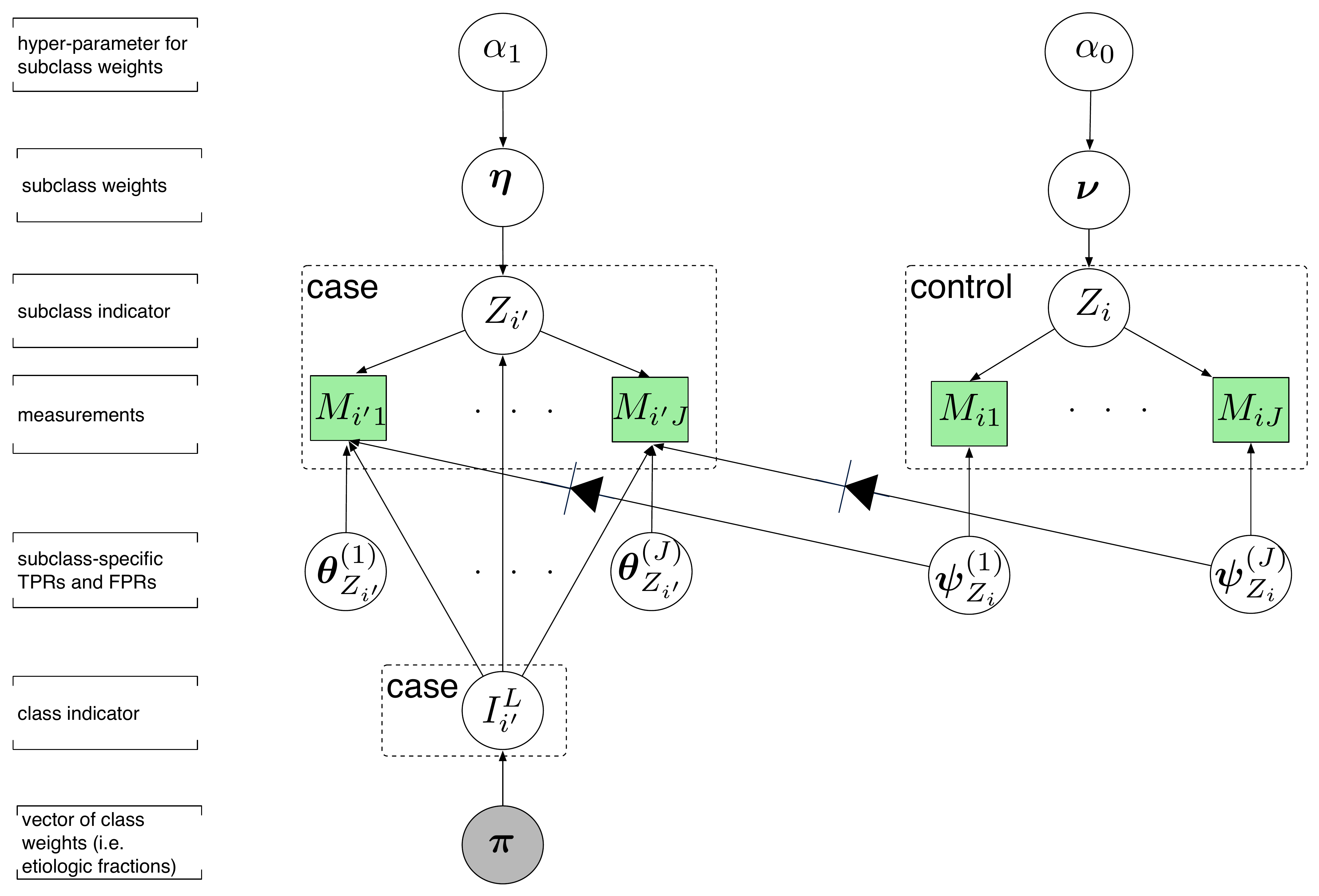}
                                                                                                                                                            \end{center}
                                                                                                                                                            \caption[Directed acyclic graph for the npLCM.]{Directed acyclic graph (DAG) for the npLCM. Quantities in circles are unknown parameters or auxiliary variables; quantities in solid squares are observables. The etiologic fraction $\bpi$ is of primary scientific interest. The solid arrows represent probabilistic relationship between the connected variables. The ``cut" valve ``A \filleddiode  B" means that when updating node A in the Gibbs sampler, we drop the likelihood terms that involve node B. }
                                                                                                                                                            \label{fig:nplcm_dag}
                                                                                                                                                            \end{figure}
                                                                                                                                                            
                                                                                                                                                            Because the false positive rate (FPR) parameters $\bPsi$ are in both the control and case likelihood (2.3) and (2.4) in the text, their posterior depend on both the control and case models. This is referred to as ``feedback" because the case model will indirectly inform $\bPsi$. If we only want the control data to inform the case model but not vice versa, we can ``cut" this source of feedback through approximate conditional updating in the Gibbs sampler  \citep{lunn2009combining}. That is, we update ${\psi}_k^{(j)}$ by $pr({\psi}_k^{(j)}\mid M_{ij}; i: Y_i=0)$ instead of Step $7$ of the Gibbs sampler (see Appendix D). It will cut the information flow from the case model to the FPR parameters $\bPsi$ and is indicated by the check-bit valves in \ref{fig:nplcm_dag}. It is desirable when certain parts of the joint model are considered not reliable to inform a subset of parameters, and can be implemented by the \verb"cut" function in \verb"WinBUGS 1.4".  Such ``cut-the-feedback" approximate Bayesian computation has both gains in computational speed and inferential robustness, and is also suggested in other contexts \citep{liu2009modularization,warren2012spatial,zigler2014uncertainty}.

                                                                                                                                                                                                                                                                                                          \FloatBarrier
                                                                                                                                                                                                                                                                                                          \section{Parameter Settings and Coverage Rates in Simulation Studies}
                                                                                                                                                                                                                                                                                                          We present the true parameter values and the empirical coverage rates in simulation studies (Section 4). 
                                                                                                                                                                                                                                                                                                          
                                                                                                                                                                                                                                                                                                          \vspace{0cm}
                                                                                                                                                                                                                                                                                                          \begin{myindentpar}{-2cm}
                                                                                                                                                                                                                                                                                                          \begin{table}[!htbp] 
                                                                                                                                                                                                                                                                                                          \begin{minipage}[t]{0.5\linewidth} 
                                                                                                                                                                                                                                                                                                          
                                                                                                                                                                                                                                                                                                          \begin{tabularx}{\textwidth}{ @{}>{}l @{\hspace{1em}} X}
                                                                                                                                                                                                                                                                                                          \underline{Scenario {\sf I} }\\[0.5em]
                                                                                                                                                                                                                                                                                                          $\bpi$  & = $(0.5,0.2,0.15,0.1,0.05)'$\\
                                                                                                                                                                                                                                                                                                          $\bTheta^T$ & = $\begin{bmatrix}
                                                                                                                                                                                                                                                                                                          0.95  & 0.9 & 0.9 & 0.9 & 0.9\\
                                                                                                                                                                                                                                                                                                          0.95  & 0.9 & 0.9 & 0.9  & 0.9\\
                                                                                                                                                                                                                                                                                                          \end{bmatrix}$\\
                                                                                                                                                                                                                                                                                                          $\bPsi^T$  & = $\begin{bmatrix}
                                                                                                                                                                                                                                                                                                          0.25 & 0.25 & 0.2 & 0.15 & 0.15\\
                                                                                                                                                                                                                                                                                                          0.2 & 0.2 & 0.25 & 0.1 & 0.1\\
                                                                                                                                                                                                                                                                                                          \end{bmatrix}$\\
                                                                                                                                                                                                                                                                                                          $\bnu$ & = $(0.5,0.5)'$\\
                                                                                                                                                                                                                                                                                                          $\bm{\eta}$ & = $(\eta_o, 1-\eta_o)'$, $0 \leq \eta_o \leq 1$
                                                                                                                                                                                                                                                                                                          \end{tabularx}
                                                                                                                                                                                                                                                                                                          
                                                                                                                                                                                                                                                                                                          \end{minipage} 
                                                                                                                                                                                                                                                                                                          \hspace{0.5cm} 
                                                                                                                                                                                                                                                                                                          \begin{minipage}[t]{0.5\linewidth} 
                                                                                                                                                                                                                                                                                                          \begin{tabularx}{\textwidth}{ @{}>{}l @{\hspace{1em}} X}
                                                                                                                                                                                                                                                                                                          \underline{Scenario {\sf II} }\\[0.5em]
                                                                                                                                                                                                                                                                                                          $\bpi$  & = $(0.5,0.2,0.15,0.1,0.05)'$\\
                                                                                                                                                                                                                                                                                                          $\bTheta^T$ & = $\begin{bmatrix}
                                                                                                                                                                                                                                                                                                          0.95  & 0.95 & 0.55 & 0.95 & 0.95\\
                                                                                                                                                                                                                                                                                                          0.95  & 0.55 & 0.95 & 0.55  & 0.55\\
                                                                                                                                                                                                                                                                                                          \end{bmatrix}$\\
                                                                                                                                                                                                                                                                                                          $\bPsi^T$  & = $\begin{bmatrix}
                                                                                                                                                                                                                                                                                                          0.4 & 0.4 & 0.05 & 0.2 & 0.2\\
                                                                                                                                                                                                                                                                                                          0.05 & 0.05 & 0.4 & 0.05 & 0.05\\
                                                                                                                                                                                                                                                                                                          \end{bmatrix}$\\
                                                                                                                                                                                                                                                                                                          $\bnu$ & = $(0.5,0.5)'$\\
                                                                                                                                                                                                                                                                                                          $\bm{\eta}$ & = $(\eta_o, 1-\eta_o)'$, $0 \leq \eta_o \leq 1$
                                                                                                                                                                                                                                                                                                            \end{tabularx}

                                                                                                                                                                                                                                                                                                          \end{minipage} 
                                                                                                                                                                                                                                                                                                          \end{table}
                                                                                                                                                                                                                                                                                                          \end{myindentpar}

                                                                                                                                                                                                                                                                                                          \begin{table}[H]
                                                                                                                                                                                                                                                                                                          \caption[]{Comparison of the actual coverage rates of $95\%$ credible intervals for each disease class estimated by results fitted to $1,000$ replication data sets.}
                                                                                                                                                                                                                                                                                                          \label{table:pie_coverage}
                                                                                                                                                                                                                                                                                                          \centering
                                                                                                                                                                                                                                                                                                          \renewcommand{\arraystretch}{1.1}
                                                                                                                                                                                                                                                                                                          \sf
                                                                                                                                                                                                                                                                                                          \begin{tabular}{cccccccccccc}
                                                                                                                                                                                                                                                                                                          \hline
                                                                                                                                                                                                                                                                                                          && & \multicolumn{9}{c}{Truth: Cases' First Subclass Weight ($\eta_o$)}\\
                                                                                                                                                                                                                                                                                                            \vspace{0.1em}
                                                                                                                                                                                                                                                                                                            &	  & Model & 0 & & 0.25 &  & 0.5 & & 0.75 & &  1  \\ 
                                                                                                                                                                                                                                                                                                            \hline
                                                                                                                                                                                                                                                                                                            &\underline{Class} & &\multicolumn{9}{c}{\multirow{1}{*}{\underline{100$\times$Coverage (Standard Error)}} }\\
                                                                                                                                                                                                                                                                                                            \multirow{10}{*}{\textbf{I}} &	\multirow{2}{*}{$A$} &  np 
                                                                                                                                                                                                                                                                                                            & \msd{ 98.5}{  0.4}  && \msd{ 99.3}{  0.3}  && \msd{ 98.5}{  0.4}  && \msd{ 98.1}{  0.4}  && \msd{ 97.8}{  0.5} \\ 
                                                                                                                                                                                                                                                                                                            &&  p 
                                                                                                                                                                                                                                                                                                            & \msd{ 97.8}{  0.5}  && \msd{ 97.6}{  0.5}  && \msd{ 98.3}{  0.4}  && \msd{ 97.7}{  0.5}  && \msd{ 96.5}{  0.6} \\ 
                                                                                                                                                                                                                                                                                                            & \multirow{2}{*}{$B$} &  np 
                                                                                                                                                                                                                                                                                                            & \msd{ 98.8}{  0.3}  && \msd{ 97.9}{  0.5}  && \msd{ 97.8}{  0.5}  && \msd{ 97.3}{  0.5}  && \msd{ 98.4}{  0.4} \\ 
                                                                                                                                                                                                                                                                                                            &&  p 
                                                                                                                                                                                                                                                                                                            & \msd{ 98.5}{  0.4}  && \msd{ 98.2}{  0.4}  && \msd{ 97.4}{  0.5}  && \msd{ 97.7}{  0.5}  && \msd{ 96.8}{  0.6} \\ 
                                                                                                                                                                                                                                                                                                            & \multirow{2}{*}{$C$} &  np 
                                                                                                                                                                                                                                                                                                            & \msd{ 96.6}{  0.6}  && \msd{ 98.5}{  0.4}  && \msd{ 97.7}{  0.5}  && \msd{ 97.7}{  0.5}  && \msd{ 94.3}{  0.7} \\ 
                                                                                                                                                                                                                                                                                                            &&  p 
                                                                                                                                                                                                                                                                                                            & \msd{ 93.0}{  0.8}  && \msd{ 96.6}{  0.6}  && \msd{ 98.6}{  0.4}  && \msd{ 97.5}{  0.5}  && \msd{ 95.1}{  0.7} \\ 
                                                                                                                                                                                                                                                                                                            & \multirow{2}{*}{$D$} &  np 
                                                                                                                                                                                                                                                                                                            & \msd{ 99.0}{  0.3}  && \msd{ 99.1}{  0.3}  && \msd{ 98.1}{  0.4}  && \msd{ 98.1}{  0.4}  && \msd{ 97.6}{  0.5} \\ 
                                                                                                                                                                                                                                                                                                            &&  p 
                                                                                                                                                                                                                                                                                                            & \msd{ 98.3}{  0.4}  && \msd{ 98.6}{  0.4}  && \msd{ 98.3}{  0.4}  && \msd{ 96.9}{  0.5}  && \msd{ 95.8}{  0.6} \\ 
                                                                                                                                                                                                                                                                                                            & \multirow{2}{*}{$E$} &  np 
                                                                                                                                                                                                                                                                                                            & \msd{ 98.1}{  0.4}  && \msd{ 98.5}{  0.4}  && \msd{ 98.2}{  0.4}  && \msd{ 98.0}{  0.4}  && \msd{ 97.4}{  0.5} \\ 
                                                                                                                                                                                                                                                                                                            &&  p 
                                                                                                                                                                                                                                                                                                            & \msd{ 98.6}{  0.4}  && \msd{ 97.1}{  0.5}  && \msd{ 96.6}{  0.6}  && \msd{ 96.3}{  0.6}  && \msd{ 95.2}{  0.7} \\ \cline{2-12} \\ 
                                                                                                                                                                                                                                                                                                            \multirow{10}{*}{\textbf{II}} &	\multirow{2}{*}{$A$} &  np 
                                                                                                                                                                                                                                                                                                            & \msd{ 95.4}{  0.7}  && \msd{ 88.4}{  1.1}  && \msd{ 88.2}{  1.1}  && \msd{ 94.0}{  0.8}  && \msd{ 98.8}{  0.3} \\ 
                                                                                                                                                                                                                                                                                                            &&  p 
                                                                                                                                                                                                                                                                                                            & \msd{ 99.6}{  0.2}  && \msd{100.0}{  0.0}  && \msd{ 96.7}{  0.6}  && \msd{ \textit{85.6}}{  1.1}  && \msd{ \textbf{72.3}}{  1.4} \\ 
                                                                                                                                                                                                                                                                                                            & \multirow{2}{*}{$B$} &  np 
                                                                                                                                                                                                                                                                                                            & \msd{\textit{ 80.4}}{  1.3}  && \msd{ \textit{84.8}}{  1.1}  && \msd{\textit{ 86.2}}{  1.1}  && \msd{ 98.3}{  0.4}  && \msd{ 98.1}{  0.4} \\ 
                                                                                                                                                                                                                                                                                                            &&  p 
                                                                                                                                                                                                                                                                                                            & \msd{  \textbf{9.9}}{  0.9}  && \msd{ \textbf{62.5}}{  1.5}  && \msd{ 92.1}{  0.9}  && \msd{ 98.9}{  0.3}  && \msd{ \textit{82.9}}{  1.2} \\ 
                                                                                                                                                                                                                                                                                                            & \multirow{2}{*}{$C$} &  np 
                                                                                                                                                                                                                                                                                                            & \msd{ 89.2}{  1.0}  && \msd{ 89.8}{  1.0}  && \msd{ 97.2}{  0.5}  && \msd{ 98.0}{  0.4}  && \msd{ \textit{84.4}}{  1.1} \\ 
                                                                                                                                                                                                                                                                                                            &&  p 
                                                                                                                                                                                                                                                                                                            & \msd{  \textbf{0.0}}{  0.0}  && \msd{  \textbf{6.1}}{  0.8}  && \msd{ 91.0}{  0.9}  && \msd{ \textbf{75.3}}{  1.4}  && \msd{  \textbf{0.0}}{  0.0} \\ 
                                                                                                                                                                                                                                                                                                            & \multirow{2}{*}{$D$} &  np 
                                                                                                                                                                                                                                                                                                            & \msd{ 93.5}{  0.8}  && \msd{ 90.7}{  0.9}  && \msd{ 95.4}{  0.7}  && \msd{ 98.0}{  0.4}  && \msd{ 95.1}{  0.7} \\ 
                                                                                                                                                                                                                                                                                                            &&  p 
                                                                                                                                                                                                                                                                                                            & \msd{ \textbf{53.3}}{  1.6}  && \msd{ 88.7}{  1.0}  && \msd{ 97.2}{  0.5}  && \msd{ 98.0}{  0.4}  && \msd{ 93.9}{  0.8} \\ 
                                                                                                                                                                                                                                                                                                            & \multirow{2}{*}{$E$} &  np 
                                                                                                                                                                                                                                                                                                            & \msd{ 95.4}{  0.7}  && \msd{ 94.7}{  0.7}  && \msd{ 96.1}{  0.6}  && \msd{ 98.5}{  0.4}  && \msd{ 96.5}{  0.6} \\ 
                                                                                                                                                                                                                                                                                                            &&  p 
                                                                                                                                                                                                                                                                                                            & \msd{ \textbf{56.1}}{  1.6}  && \msd{ 92.1}{  0.9}  && \msd{ 97.8}{  0.5}  && \msd{ 98.2}{  0.4}  && \msd{ 92.0}{  0.9} \\
                                                                                                                                                                                                                                                                                                            \hline\hline
                                                                                                                                                                                                                                                                                                            
                                                                                                                                                                                                                                                                                                            \end{tabular}
                                                                                                                                                                                                                                                                                                            \end{table}
                                                                                                                                                                                                                                                                                                            
                                                                                                                                                                                                                                                                                                            \FloatBarrier
                                                                                                                                                                                                                                                                                                            \newpage
                                                                                                                                                                                                                                                                                                            \section{For Section 5: Analysis of PERCH Data}
                                                                                                                                                                                                                                                                                                            \label{appendix:data.analysis}
                                                                                                                                                                                                                                                                                                            \textbf{Full Pathogen Names and Abbreviations}: \\
                                                                                                                                                                                                                                                                                                            (1).{\small \sf HINF}- \textit{Haemophilus Influenzae}; (2). {\small \sf ADENO} -Adenovirus; (3). {\small \sf HMPV-A/B} - Human Metapneumovirus Type A or B; (4). {\small \sf PARA-1} - Parainfluenza Type 1 Virus; (5). {\small \sf RHINO} - Rhinovirus; (6). {\small \sf RSV} - Respiratory Syncytial Virus Type A or B.
                                                                                                                                                                                                                                                                                                            
                                                                                                                                                                                                                                                                                                            \begin{figure}[!htbp]
                                                                                                                                                                                                                                                                                                            \begin{center}
                                                                                                                                                                                                                                                                                                            \includegraphics[width=\linewidth]{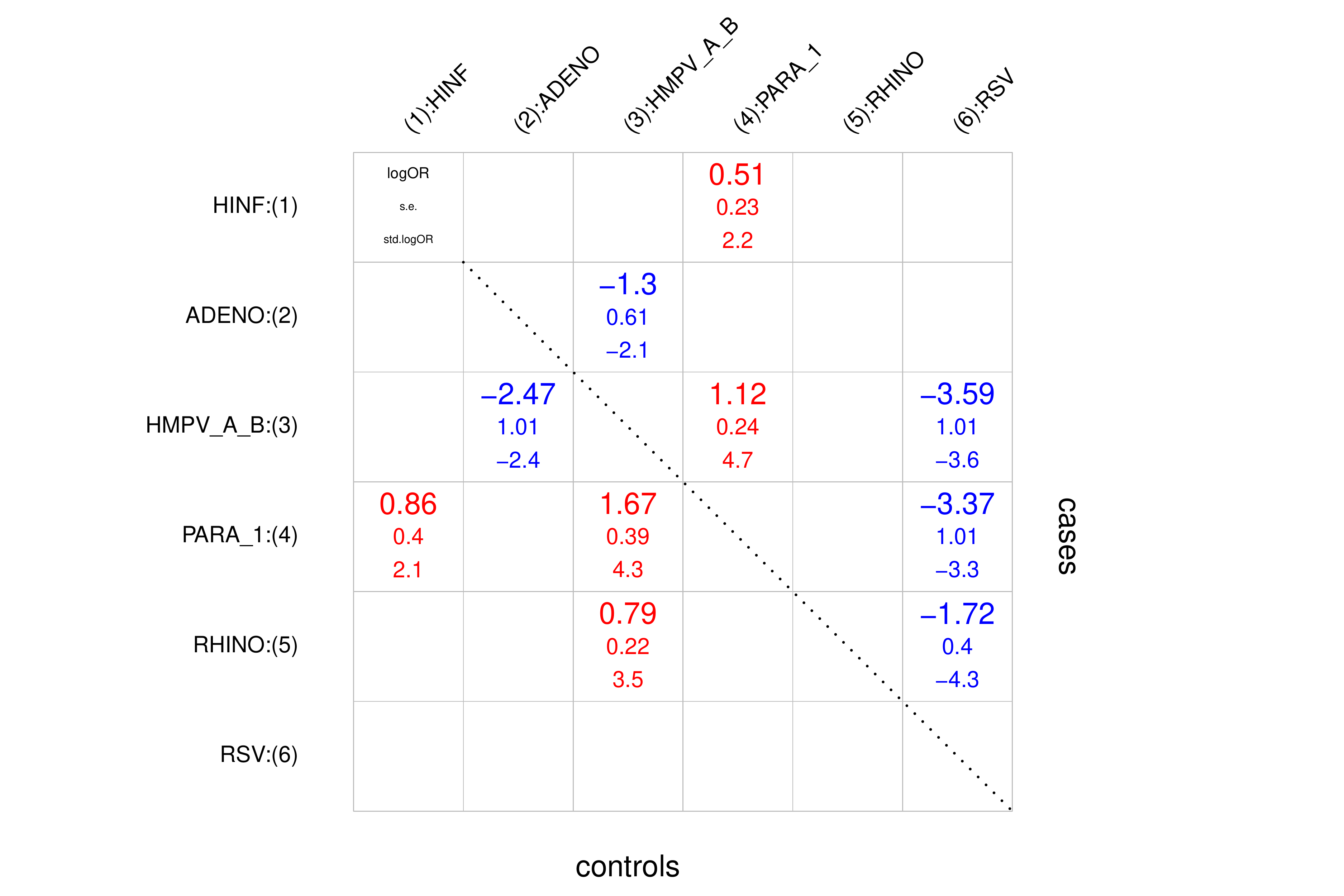}
                                                                                                                                                                                                                                                                                                            \end{center}
                                                                                                                                                                                                                                                                                                            \caption{Matrix of significant pairwise log odds ratios (LOR) for cases (upper) and controls (lower). LOR is at the top of the cell. Below it, its standard error is in smaller type, using the same color as the LOR. Then the estimate is divided by its standard error. We put the actual value when the Z-statistics has an absolute value greater than $2$; a plus (red) or minus (blue) if between $1$ and $2$; blank otherwise. }
                                                                                                                                                                                                                                                                                                            \label{fig:significant_LOR}
                                                                                                                                                                                                                                                                                                            \end{figure}
                                                                                                                                                                                                                                                                                                            
                                                                                                                                                                                                                                                                                                            \begin{figure}[H]
                                                                                                                                                                                                                                                                                                            \centering
                                                                                                                                                                                                                                                                                                            \includegraphics[width=\linewidth]{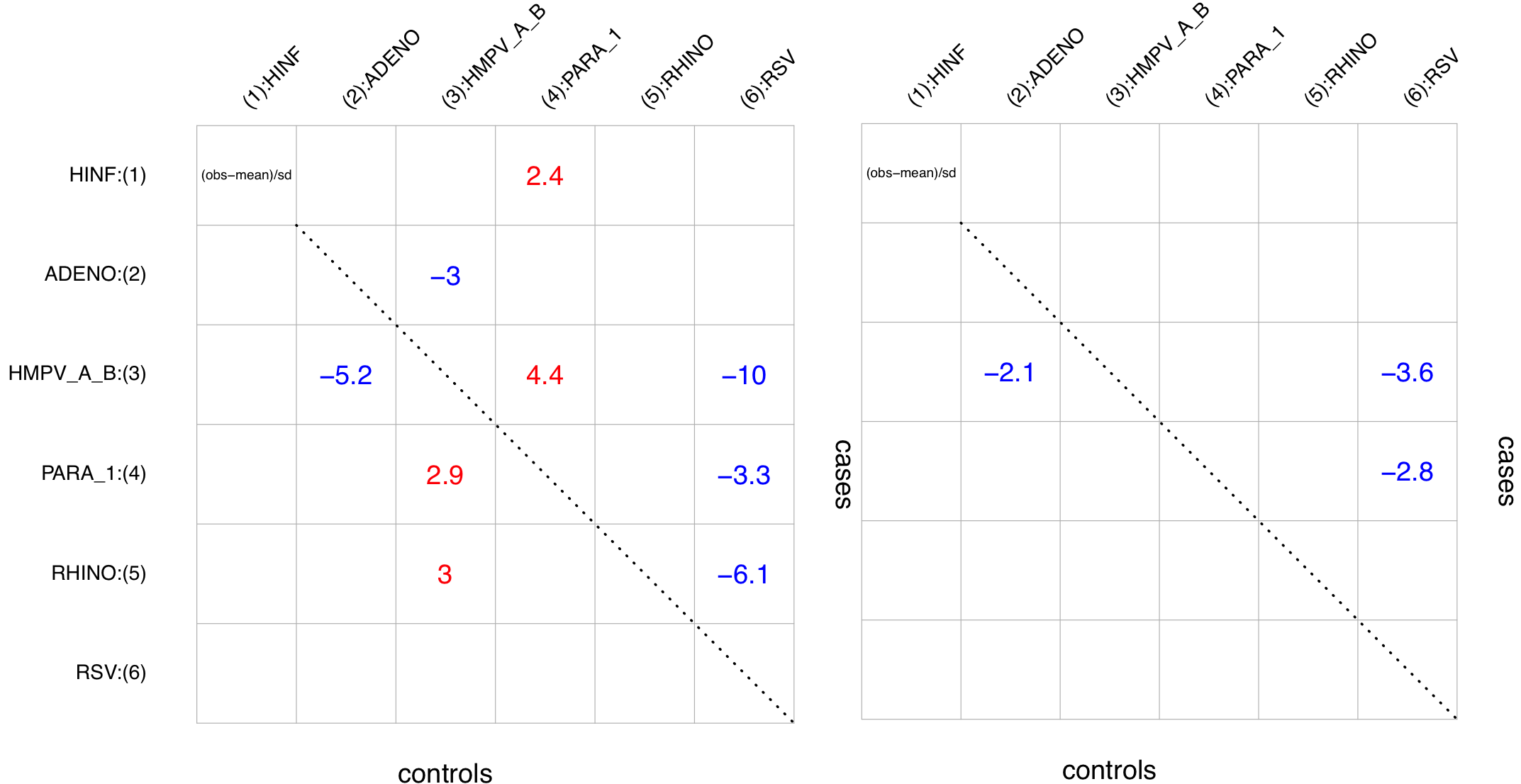}
                                                                                                                                                                                                                                                                                                            \caption{Posterior predictive checking for pairwise odds ratios separately for cases (upper triangle) and controls (lower triangle) with expert priors on true positive rates. \textit{Left}: pLCM; \textit{Right}: npLCM. Each entry is a standardized log odds ratio difference (SLORD): the observed log odds ratio for a pair of measurements minus the mean LOR for the posterior predictive distribution divided by the standard deviation of the posterior predictive distribution. The first significant digit of absolute SLORs are shown in red for positive and blue for negative values, and only those greater than 2 are shown. On average, for a well fitting model, we expect $0.05\times\binom{6}{2} \times 2\approx 1.5 (\pm 2.4)$ non-blank cells in cases and controls, respectively.}
                                                                                                                                                                                                                                                                                                            \label{fig:checking.SLORD}
                                                                                                                                                                                                                                                                                                            \end{figure}

                                                                                                                                                                                                                                                                                                            \newpage
                                                                                                                                                                                                                                                                                                            \begin{landscape}
                                                                                                                                                                                                                                                                                                            \begin{figure}[!p]
                                                                                                                                                                                                                                                                                                            \centering     
                                                                                                                                                                                                                                                                                                            \includegraphics[width=\linewidth]{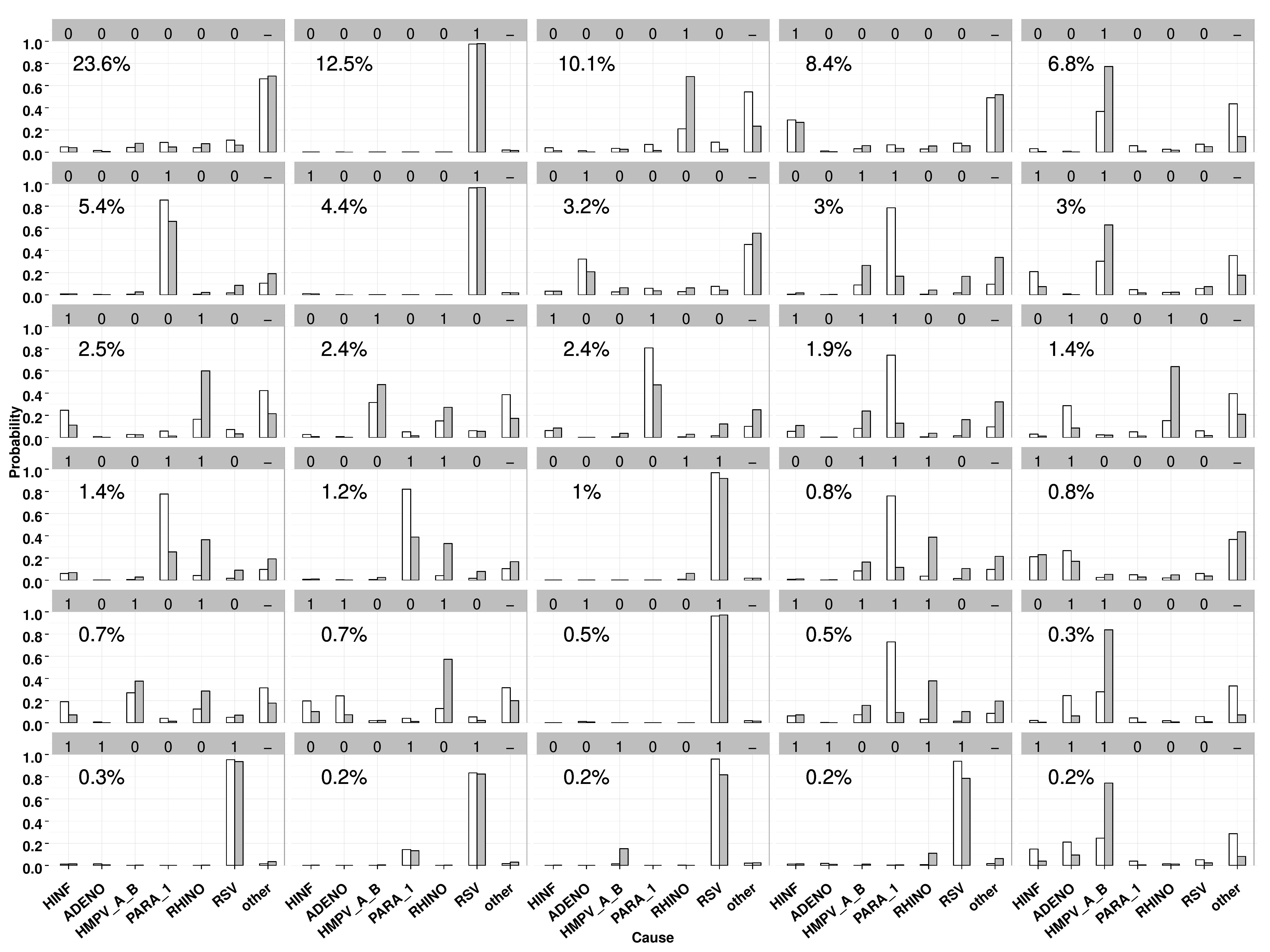}
                                                                                                                                                                                                                                                                                                            \caption{Each panel shows the individual etiology distribution estimated by the empirical distribution of MCMC samples of the disease class indicator. The height of a bar represents the probability of a case caused by each of the $7$ causes labelled on the x-axis. For each cause, paired bars compare the estimates from the pLCM (left) and the npLCM (right); the binary codes at the top represent the NPPCR data with its observed frequency marked beneath (no measurements on ``other" causes hence left as ``-"). $30$ most common patterns are shown here ordered by their observed frequencies.}
                                                                                                                                                                                                                                                                                                            \label{fig:app.ind.pred}
                                                                                                                                                                                                                                                                                                            \end{figure}
                                                                                                                                                                                                                                                                                                            \end{landscape}
                                                                                                                                                                                                                                                                                                            

                                                                                                                                                                                                                                                                                                            \end{document}